\documentclass[aps,prd,preprint,showpacs]{revtex4-1}
\usepackage{graphicx}
\usepackage{amsmath,amssymb}
\usepackage{subfigure}
\usepackage[T1]{fontenc} 
\usepackage[above,below]{placeins}
\usepackage{multirow}
\usepackage{longtable}
\usepackage{array}
\allowdisplaybreaks[2]
\begin{document}
\title{Pairwise mode entanglement in Schwinger production of particle-antiparticle pairs in an electric field}
\author{Yujie Li}
\affiliation{Center for Field Theory and Particle Physics, Department of Physics  \&  State Key Laboratory of Surface Physics,   Fudan University,\\ Shanghai 200433, China}
\author{Yue Dai}
\affiliation{Center for Field Theory and Particle Physics, Department of Physics  \&  State Key Laboratory of Surface Physics,   Fudan University,\\ Shanghai 200433, China}
\author{Yu Shi }  \email{yushi@fudan.edu.cn}
\affiliation{Center for Field Theory and Particle Physics, Department of Physics  \&  State Key Laboratory of Surface Physics,   Fudan University,\\ Shanghai 200433, China}
\affiliation{Collaborative Innovation Center of Advanced Microstructures, Fudan University, \\Shanghai 200433, China}

\begin{abstract}
Quantum entanglement is the characteristic quantum correlation. Here we use this concept to analyze the quantum entanglement generated by Schwinger production of particle-antiparticle pairs in an electric field, as well as the change of mode entanglement as a consequence of the electric field effect on an entangled pair of particles. The system is partitioned  by using momentum modes. Various kinds of pairwise mode entanglement are calculated as functions of the electric field. Both constant and pulsed electric fields are considered.  The use of entanglement exposes information beyond that  in particle number distributions.
\end{abstract}


\maketitle

\section{Introduction}

With some early predecessors~\cite{sauter,heisenberg},
Schwinger predicted  that in a strong electromagnetic field, the vacuum becomes unstable and decays to electron-positron pairs~\cite{Schwinger}. This is known as Schwinger effect or Schwinger production of  electron-positron pairs. As a nonperturbative  prediction of quantum electrodynamics, it has not yet  been observed experimentally because of the low rate $\sim \exp [-\pi m_e^2/(eE)]$, where $E$ is the electric field, $m_e$ is the mass of the electron.  It has been a long pursuit in the fields of strong lasers~\cite{rmp} and heavy-ion collisions~\cite{cooper}, and there have been various proposals to improve the  detection~\cite{gies,tajima,dunne1,Schutzhold,dunne2,Piazza,Bulanov,
Monin,Dumlu1,hebenstreit,gourdeau,gonoskov,Blinne,Jansen,Wollert}. Recently, Schwinger effect has even been generalized to particle-antiparticle pair production in the supersymmetric Yang-Mills theory~\cite{semenoff,sonner}.

As a quantum effect in a classical background, converting vacuum fluctuation into particles,  Schwinger effect bears similarities with the Unruh effect as observed by an accelerating detector and the  Hawking effect of a black hole~\cite{brout}, as well as  the inflationary cosmological perturbation~\cite{martin}. In this connection, there have been various calculations of the Bogoliubov coefficients in the Schwinger effect~\cite{kim}, which relates the ``out'' state with the ``in'' state and underly the mechanism of particle creations.

On the other hand, in recent years, there has been significant development in quantum information science, not only in practical aspects, but also in fundamental aspects, for example, in deepening our understanding of correlation and entanglement in quantum states. It was proposed that in field theory, the proper approach of characterizing the quantum entanglement is to use modes,  designated by the coordinates or momenta, as the subsystems~\cite{shi}. Some concepts developed in quantum information theory have been applied to quantum states near a black hole~\cite{fuentes1,em1,em2,eduardo,dai1,dai2} and in an expanding universe~\cite{ball,Fuentes3,Moradi,li}.
This effort has also been extended to  Schwinger effect, and  the  entanglement between one Dirac mode and the rest of the system was calculated~\cite{Ebadi}. Such entanglement was also studied  in the more general framework of QED  by using  a nonperturbative expression for the density operators~\cite{gavrilov}. However, the information captured by the kind of entanglement as studied in these two papers is limited.

More interesting and accessible is the pairwise entanglement, in analogy with the fact that two-body correlations in a many-body system, rather than the correlation between one-body and all the other bodies, are the usual quantities  studied in physics. Here pairwise entanglement refers to the entanglement between two parts A and B in the  reduced density matrix $\rho_{AB}$, obtained by tracing out other parts. Note that any pair is usually in a mixed state, obtained by tracing out other subsystems, therefore the characterization of pairwise entanglement is more difficult than the entanglement between one subsystem and its complement in the system in a pure state, as quantified by the von Neumann entropy of the reduced density matrix. Furthermore, there is no simple relation between pairwise entanglement and the entanglement between one part and all the rest of the system. For example, in the famous  tripartite GHZ state $\frac{1}{\sqrt{2}} (|000\rangle +|111\rangle)$,  the entanglement between any part and its complement is maximal, while any pairwise entanglement vanishes because tracing out the third part yields an incoherent mixed state. Furthermore, in general, the pairwise entanglement between different pairs can be very different. In this paper, we shall study the pairwise entanglement and correlations in Schwinger  effect by using two measures of entanglement and correlation, namely, mutual information  and logarithmic negativity.

Consider a system composed of two subsystems A and B. Suppose the system, which could itself be a subsystem of a larger system, is described by a density matrix $\rho_{AB}$, which could be a pure state or a mixed state, the reduced density matrix of A is  $\rho_A \equiv Tr_B(\rho_{AB})$, the the reduced density matrix of B is  $\rho_B \equiv Tr_A(\rho_{AB})$. Then the mutual information is defined as~\cite{nielsen}
\begin{equation}
I(\rho_{AB}) = S(\rho_A) +S(\rho_B) - S(\rho _{AB}), \end{equation}
where $S(\rho)\equiv -Tr(\rho\log_2\rho)$ is the von Neumann entropy of density matrix $\rho$.    For given $\rho_A$ and $\rho_B$, it is maximized when the $\rho _{AB}$ is a pure state and thus $S(\rho _{AB})=0$.  Moreover, if  $\rho _{AB}$ is a pure state, then $S(\rho_A)=S(\rho_B) $. If $\rho _{AB}=\rho_A\otimes \rho_B$, $I(\rho _{AB})=0$. Hence $I(\rho_{AB})$ is a kind of  ``distance'' from the product state, and it is a kind of  total correlation between A and B, including both quantum entanglement and classical correlation.

The logarithmic negativity is defined as~\cite{vidal}
\begin{equation}
N(\rho_{AB} ) \equiv \log _2 \| \rho_{AB}^{T_A} \| , \end{equation}
where $ \| \rho_{AB} ^{T_A} \|$ is the sum of the absolute values of the eigenvalues of  the partial transpose $\rho ^{T_A}$  of the original density matrix $\rho_{AB}$  with respect to A  subsystem. The same value of  $N(\rho_{AB} )$ is obtained if the partial transpose is with respect to B. It has been shown that $N(\rho_{AB} )$ is a measure of the quantum entanglement between A and B in a general density matrix $\rho_{AB} $.

The vacuum in the absence of an electric field is not the vacuum in the presence of an electric field. Consider the sector of a pair of modes with opposite momenta, the vacuum in the absence of an electric field is a superposition of different occupations of the the two modes in the presence of the electric field. In Sec.~\ref{const}, we calculate the entanglement between these two modes, which is compared with the previous results. On the other hand, the state of two entangled particles in the absence of the electric field becomes a complicated superposed state of these two modes and those  with the opposite momenta. In Sec.~\ref{ent}, we study various kinds of pairwise mode entanglement in this state. In Sec.~\ref{pulse}, we study the above two problems in a pulsed electric field. A summary and discussion are made in Sec.~\ref{summary}.

\section{Entanglement creation in the constant electric field \label{const}  }

A strong electric field can destablize the vacuum to decay into electrically charged particle-antiparticle pairs. Consider a fermion field $\Psi(x)$ with  mass $m$, charge $e$  in an electric field $E$ along $z$ direction, satisfying the Dirac equation in the four dimensional Minkowski space,
\begin{equation} (i\gamma ^\mu \partial _\mu  - e\gamma ^\mu A_\mu - m)\Psi (x) = 0 , \end{equation}
where $A_\mu  = (0,0,0, - E_0 t)$, $\gamma ^\mu $'s  are Dirac matrices.

Suppose $\{ u^{\mathrm{in}}, v^{\mathrm{in}}\} $ and $\{ u^{\mathrm{out}}, v^{\mathrm{out}}\} $ are two complete sets of mode functions of Dirac equation, where  ``$\mathrm{in}$'' and ``$\mathrm{out}$''  states are those in the absence and the presence of the  electric field,  corresponding  to $t_{\mathrm{in}}=-\infty$ and  $t_{\mathrm{out} }=+ \infty$,  respectively.   They are related through the Bogoliubov  transformation. Hence the Dirac field operator can be expanded  as
\begin{equation}
\begin{split}
\Psi (x) &= \sum\limits_{\sigma,\mathbf{k}}[a_{\sigma,\mathbf{k}}^{\mathrm{in}}u_{\sigma,\mathbf{k}}^{\mathrm{in}}(x) + b_{\sigma,\mathbf{k}}^{{\mathrm{in}}\dag} v_{\sigma,\mathbf{k}}^{\mathrm{in}}(x)]\\
         &= \sum\limits_{\sigma,\mathbf{k}}[a_{\sigma,\mathbf{k}}^{\mathrm{out}}u_{\sigma,\mathbf{k}}^{\mathrm{out}}(x) + b_{\sigma,\mathbf{k}}^{{\mathrm{out}}\dag}
v_{\sigma,\mathbf{k}}^{\mathrm{out}}(x)] ,
\end{split}
\end{equation}
where $a_{\sigma,\mathbf{k}}$ and $b_{\sigma,\mathbf{k}}$ are annihilation operators of particle and antiparticle, $\sigma=\uparrow ,\downarrow$ denotes the spin,  $\mathbf{k}$ denotes the momentum. It has been well known that (see, for example, Ref.~\cite{kim,Ebadi}) the Bogoliubov transformations between $\mathrm{in}$ and $\mathrm{out}$ operators are
\begin{equation} a_{\sigma,\mathbf{k}}^{\mathrm{in}} = \alpha _{\mathbf{k}}^* a_{\sigma,\mathbf{k}}^{\mathrm{out}} + \beta _{\mathbf{k}}^* b_{-\sigma,-\mathbf{k}}^{\mathrm{out}\dag },
\end{equation}
\begin{equation}  b_{\sigma,\mathbf{k}}^{\mathrm{in}} = \alpha _{\mathbf{k}}^* b_{\sigma,\mathbf{k}}^{\mathrm{out}} - \beta _{\mathbf{k}}^* a_{-\sigma,-\mathbf{k}}^{\mathrm{out}\dag } ,  \end{equation}
where
$\alpha _{ \mathbf{k}}    = \sqrt {\frac{\mu }{\pi }} \Gamma (\frac{i\mu }{2})\sinh (\frac{\pi \mu }{2})e^{ - \frac{\pi \mu }{4}},$ $\beta _{\mathbf{k}} =  e^{ - \frac{\pi \mu }{2}}$ ,   hence
\begin{equation}  |\alpha _\mathbf{k}|^2 = 1 - e^{ - \pi \mu }\ \ ,\ \ | \beta _\mathbf{k} |^2 = e^{ - \pi \mu }.   \label{alphabeta} \end{equation}
with
\begin{equation}
 \mu  = \frac{m^2 + k_ \bot ^2}{e E_0}\ \ ,\ \  k_ \bot ^2 = k_x^2 + k_y^2 .  \end{equation}

As usual~\cite{brout,martin,kim,shi, fuentes1,em1,em2,eduardo,dai1,dai2,ball,Fuentes3,Moradi,li,Ebadi},
the $\mathrm{in}$ vacuum can be expanded in terms of the $\mathrm{out}$ states of modes $\mathbf{k}$ and  $-\mathbf{k}$,
\begin{equation}  | 0_\mathbf{k},0_{ - \mathbf{k}} \rangle ^{\mathrm{in}} = x_0 | 0_\mathbf{k},0_{ - \mathbf{k}} \rangle ^{\mathrm{out}} + x_1 |  \uparrow _\mathbf{k}, \downarrow _{ -\mathbf{k}}\rangle ^{\mathrm{out}} + x_2| \downarrow _\mathbf{k},\uparrow_{-\mathbf{k}} \rangle ^{\mathrm{out}}+ x_3 | \uparrow  \downarrow _\mathbf{k}, \uparrow  \downarrow _{ -\mathbf{k}}\rangle ^{\mathrm{out}}, \label{vacuum}
  \end{equation}
where $|\phi_\mathbf{k},\psi_{-\mathbf{k}}\rangle \equiv  |\phi\rangle_\mathbf{k} |\psi\rangle_{-\mathbf{k}}$,  $| 0\rangle_\mathbf{k}$ means that there is no particle with momentum $\mathbf{k}$,   $|0_\mathbf{k}\rangle^{\mathrm{in}}$ and $|0_\mathbf{k}\rangle^{\mathrm{out}}$ denote $\mathrm{in}$ vacuum state and $\mathrm{out}$  vacuum state respectively, $|\sigma\rangle_\mathbf{k}$ represents that there is particle of spin $\sigma$ with momentum $\mathbf{k}$, $|\uparrow  \downarrow \rangle_\mathbf{k} $ represents that there are two particles of spins $\uparrow$  and $\downarrow$  with momentum $\mathbf{k}$, $x_0$, $x_1$, $x_2$ and $x_3$ are expansion coefficients. We treat  the momentum modes as the subsystems~\cite{shi}, with each momentum mode $\mathbf{k}$ living in a four-dimensional Hilbert space with the basis states $|0\rangle_\mathbf{k}$, $|\uparrow\rangle_\mathbf{k}$,  $|\downarrow\rangle_\mathbf{k}$, $|\uparrow \downarrow\rangle_\mathbf{k}$. In this way, the spin correlations between different momentum modes are investigated.

From $a_{\sigma,\mathbf{k}}^{\mathrm{in}}| 0_\mathbf{k},0_{ -\mathbf{k}}\rangle ^{\mathrm{in}} = 0$,  one obtains
\begin{equation}\begin{array}{rl}
| 0_\mathbf{k},0_{ -\mathbf{k}}\rangle ^{\mathrm {in}} =& | \alpha _\mathbf{k}|^2| 0_\mathbf{k},0_{ -\mathbf{k}}\rangle ^{\mathrm{out}} - \alpha _\mathbf{k} \beta _\mathbf{k}^*| \uparrow _\mathbf{k}, \downarrow _{ -\mathbf{k}}\rangle ^{\mathrm{out}} - \alpha _\mathbf{k} \beta _\mathbf{k}^* | \downarrow _\mathbf{k},\uparrow _{ - \mathbf{k}}\rangle ^{\mathrm{out}}\nonumber\\
&+ \frac{\alpha _\mathbf{k} \beta _\mathbf{k}^{*2}}{\alpha _\mathbf{k}^*}| \uparrow \downarrow _\mathbf{k} , \uparrow \downarrow _{- \mathbf{k}}\rangle ^{\mathrm{out}}.
\end{array}\end{equation}
The density matrix is $\rho _{\mathbf{k}, - \mathbf{k}}=| 0_\mathbf{k},0_{ -\mathbf{k}}\rangle ^{\mathrm {in}}\langle 0_\mathbf{k},0_{ -\mathbf{k}}|$. Hence one obtains its partial transpose
\begin{equation}\begin{array}{rl}
\rho _{\mathbf{k}, - \mathbf{k}}^{T_A} =& \frac{1}{2}[| \alpha _\mathbf{k}|^4 | 0_\mathbf{k} , 0_{ -\mathbf{k}}\rangle \langle 0_\mathbf{k},0_{ - \mathbf{k}}| - 2| \alpha _\mathbf{k}|^2\alpha_\mathbf{k}^* \beta_\mathbf{k}|\uparrow_\mathbf{k},0_{ - \mathbf{k}}\rangle \langle 0_\mathbf{k}, \downarrow _{ - \mathbf{k}}|\nonumber\\
&- 2|\alpha _\mathbf{k}|^2 \alpha _\mathbf{k}^*\beta _\mathbf{k}|\downarrow_\mathbf{k},0_{ - \mathbf{k}}\rangle \langle 0_\mathbf{k}, \uparrow _{ - \mathbf{k}} | + 2\alpha _\mathbf{k}^{*2}\beta _\mathbf{k}^2| \uparrow \downarrow_\mathbf{k},0_{ - \mathbf{k}}\rangle \langle 0_\mathbf{k},\uparrow  \downarrow _{ - \mathbf{k}}|\nonumber\\
&+ |\alpha _\mathbf{k}|^2 | \beta _\mathbf{k}|^2 | \uparrow _\mathbf{k}, \downarrow _{ - \mathbf{k}}\rangle \langle \uparrow _\mathbf{k}, \downarrow _{ - \mathbf{k}}| + 2| \alpha _\mathbf{k}|^2 | \beta _\mathbf{k}|^2 | \downarrow_\mathbf{k}, \downarrow _{ - \mathbf{k}}\rangle \langle \uparrow_\mathbf{k}, \uparrow _{ - \mathbf{k}}| \nonumber\\
&- 2\alpha _\mathbf{k}^*|\beta _\mathbf{k}|^2 \beta _\mathbf{k} | \uparrow \downarrow_\mathbf{k}, \downarrow _{ - \mathbf{k}}\rangle \langle \uparrow_\mathbf{k},\uparrow \downarrow _{ - \mathbf{k}}| +|\alpha _\mathbf{k}|^2 | \beta _\mathbf{k} |^2 | \downarrow _\mathbf{k}, \uparrow _{ - \mathbf{k}}\rangle \langle \downarrow _\mathbf{k}, \uparrow _{ - \mathbf{k}} |\nonumber\\
& - 2\alpha _\mathbf{k}^*| \beta _\mathbf{k}|^2 \beta _\mathbf{k} | \uparrow \downarrow _\mathbf{k}, \uparrow _{ - \mathbf{k}} \rangle \langle  \downarrow_\mathbf{k},\uparrow  \downarrow_{ - \mathbf{k}} | +  |  \beta _\mathbf{k} |^4 |  \uparrow  \downarrow _\mathbf{k}, \uparrow \downarrow _{ - \mathbf{k}} \rangle  \langle \uparrow \downarrow_\mathbf{k}, \uparrow  \downarrow_{ - \mathbf{k}}|]   \\
&+ \rm{H}\rm{.c},
\end{array}\end{equation}
which is a $16\times 16$ matrix.  Its  eigenvalues are found to be $\lambda_1 = | \alpha _\mathbf{k} |^4$, $\lambda_2 = | \beta_\mathbf{k} |^4$, $\lambda_{3}=\lambda_4=\lambda_5=\lambda_6 = | \alpha_\mathbf{k} |^2 | \beta_\mathbf{k} |^2 $, $\lambda_{7}=\lambda_8 = -| \alpha_\mathbf{k} |^2 | \beta_\mathbf{k} |^2$,    $\lambda_9=\lambda_{10} = | \alpha_\mathbf{k} |^3 | \beta_\mathbf{k} |$, $\lambda_{11} =\lambda_{12} = -| \alpha_\mathbf{k} |^3 | \beta_\mathbf{k} |$, $\lambda_{13}=\lambda_{14} =| \alpha_\mathbf{k} | | \beta_\mathbf{k} |^3$, $\lambda_{15} =\lambda_{16} = -| \alpha_\mathbf{k} | | \beta_\mathbf{k} |^3.$
Thus the logarithmic negativity is
\begin{equation}
N( \rho_{\mathbf{k}, - \mathbf{k}} ) = 4 \log _2 ( | \alpha _\mathbf{k} | + | \beta _\mathbf{k} |),  \end{equation}
which reaches its maximum $2$ if $|\alpha_\mathbf{k}|^2=|\beta_\mathbf{k}|^2=\frac{1}{2}$, which is satisfied when $\mu = \ln 2/\pi$, that is, $E_0= \frac{\pi(k_\bot +m^2)}{e \ln 2 }$. The maximum is $2$ because the number of the basis states for each mode is $4$.

\begin{figure}[htb]
\centering
\includegraphics[width=0.6\textwidth]{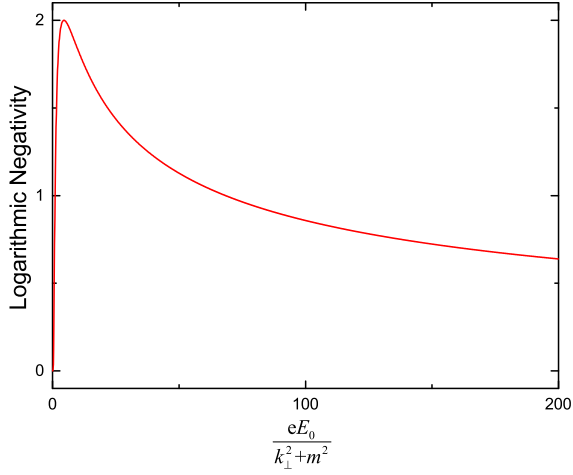}
\caption{The logarithmic negativity $N( \rho_{\mathbf{k}, - \mathbf{k}} )$ of pair of modes $\mathbf{k}$ and $ - \mathbf{k}$  as a function of the dimensionless parameter $\frac{eE_0}{k_\perp^2+m^2}$, where  $E_0$  is the strength of the constant electric field.   \label{fig1} }
\end{figure}

Fig.~\ref{fig1} shows the dependence on $1/\mu\equiv eE_0/(k_\bot +m^2)$   of the logarithmic negativity, quantifying the entanglement between modes $\mathbf{k}$ and $-\mathbf{k}$.  Therefore, it can be seen that, with the increase of $E_0$ and thus $1/\mu$, the logarithmic negativity increases to the maximum $2$ at $E_0= \frac{\pi(k_\bot +m^2)}{e \ln 2 }$,    afterwards the logarithmic negativity asymptotically  decreases to zero  with  the electric field approaching infinity.

The dependence of the entanglement on the electric field can be understood. Expressed as the Bogoliubov coefficients, an in particle creation operator  is a superposition of out particle creation and antiparticle annihilation. With the increase of $E_0$ from $0$, the degree of superposition increases, and  becomes equal probability  at a certain value $E_0$, consequently the entanglement between the particle and antiparticle modes becomes maximal. With further increase of $E_0$, the antiparticle annihilation overweighs particle creation in the superposition,  consequently the entanglement decreases towards zero. Hence the entanglement between the modes $\mathbf{k}$ and $-\mathbf{k}$ provides an interesting characterization of the superposition of out particle creation and antiparticle annihilation, and that of out antiparticle creation and particle annihilation, complements the usual characterization in terms of particle number distribution.

The result is in consistency with the previous calculation of the von Neumann entropy~\cite{Ebadi}.  In that calculation,  however, a mode is designated  in terms of momentum plus  spin, hence the system was partitioned to four subsystems there. In practice, it is more difficult to  measure and follow the  spin degree of freedom. So our result, based on designating a mode only in terms of the momentum, is more convenient. We use the negativity as the measure of entanglement, rather than von Neumann entropy because the negativity will be used in the rest of this work, which is mainly about mixed states, for which von Neumann entropy is not an entanglement measure. Furthermore, we do not need to calculate the  mutual information, which is generically contributed by both quantum entanglement and classical correlation but reduces to $2S(\rho_A)$ in the present case, as $S(\rho_{AB})=0$ in a pure state.

\section{Various kinds of mode entanglement  in an initially entangled particle pair  \label{ent} }

In this section, we study how the mode entanglement in a preexisting two-particle entangled state is affected by a constant electric field. We consider the two entangled particles with momenta $\mathbf{p}$ and $\mathbf{q}$ in the absence of electric field,
\begin{equation}
\varepsilon |\uparrow\rangle_{\mathbf{p}}|\downarrow\rangle_{\mathbf{q}}
+\sqrt{1-\varepsilon ^2} |\downarrow\rangle_{\mathbf{p}}|\uparrow\rangle_{\mathbf{q}},
\label{initialstate} \end{equation}
where $\varepsilon$ is a coefficient.
In terms of in modes, the state of the system  can be written as
\begin{equation}
\begin{array}{rl} |\Phi_{\mathbf{p},\mathbf{q},-\mathbf{p},-\mathbf{q}}\rangle ^{\mathrm{in}}& = \varepsilon | \uparrow _\mathbf{p},0_{ - \mathbf{p}} \rangle ^{ \mathrm{in}} | \downarrow _\mathbf{q},0_{ - \mathbf{q}} \rangle ^{ \mathrm{in}}  + \sqrt {1 - \varepsilon ^2}  | \downarrow _\mathbf{p},0_{ - \mathbf{p}} \rangle^{ \mathrm{in}} | \uparrow _\mathbf{q},0_{ - \mathbf{q}} \rangle^{\mathrm{in}}\\
&=( \varepsilon | \uparrow _\mathbf{p}, \downarrow _\mathbf{q}\rangle ^{ \mathrm{in}}+\sqrt {1 - \varepsilon ^2} | \downarrow _\mathbf{p}, \uparrow _\mathbf{q}\rangle ^{ \mathrm{in}} ) | 0_{ - \mathbf{p}} ,0_{ - \mathbf{q}} \rangle ^{ \mathrm{in}}.
\end{array}
\label{initialent} \end{equation}

Now suppose  an electric field is applied. One can obtain
\begin{equation}  |\uparrow _\mathbf{k}, 0_{ - \mathbf{k}}\rangle ^{\mathrm{in}} = a_{ \uparrow ,\mathbf{k}}^{\mathrm{in}\dag }| 0_\mathbf{k},0_{ - \mathbf{k}}\rangle ^{\mathrm{in}} = \alpha _\mathbf{k} | \uparrow _\mathbf{k}, 0_{ - \mathbf{k}} \rangle ^{\mathrm{out}} - \frac{\alpha _\mathbf{k} \beta _\mathbf{k}^*}{\alpha _\mathbf{k}^*}| \uparrow  \downarrow _\mathbf{k}, \uparrow _{ - \mathbf{k}}\rangle ^{\mathrm{out}}  , \end{equation}
\begin{equation}  | \downarrow _\mathbf{k},0_{ - \mathbf{k}}\rangle ^{\mathrm{in}} = a_{ \downarrow ,\mathbf{k}}^{\mathrm{in}\dag }| 0_\mathbf{k},0_{ - \mathbf{k}} \rangle ^{\mathrm{in}} = \alpha _\mathbf{k} | \downarrow _\mathbf{k},0_{ -\mathbf{k}}\rangle ^{\mathrm{out}} + \frac{\alpha _\mathbf{k} \beta _\mathbf{k}^*}{\alpha _\mathbf{k}^*}| \uparrow  \downarrow _\mathbf{k}, \downarrow _{ - \mathbf{k}}\rangle ^{\mathrm{out}} .   \end{equation}
Therefore
\begin{equation}
\begin{array}{rl}
 | \Phi _{\mathbf{p},\mathbf{q},-\mathbf{p},-\mathbf{q}} \rangle^{\mathrm{in}}= & \varepsilon \alpha_\mathbf{p}\alpha_\mathbf{q} | \uparrow_\mathbf{p},0_{-\mathbf{p}}\rangle ^{\mathrm{out}}| \downarrow_\mathbf{q},0_{-\mathbf{q}}\rangle^{\mathrm{out}} \\
&+ \varepsilon \frac{\alpha_\mathbf{p} \alpha_ \mathbf{q} \beta _\mathbf{q}^* }{ \alpha _\mathbf{q}^*}| \uparrow _\mathbf{p},0_{-\mathbf{p}}\rangle ^{\mathrm{out}}| \uparrow  \downarrow_\mathbf{q}, \downarrow_{-\mathbf{q}}\rangle^{\mathrm{out}} \\
& - \varepsilon \frac{ \alpha _\mathbf{q} \alpha _\mathbf{p} \beta _\mathbf{p}^*} {\alpha _\mathbf{p}^*} | \uparrow \downarrow_\mathbf{p}, \uparrow _{-\mathbf{p}}\rangle^{\mathrm{out}}| \downarrow_\mathbf{q},0_{-\mathbf{q}}\rangle^{\mathrm{out}} \\
& - \varepsilon \frac{\alpha_\mathbf{p} \alpha_\mathbf{q} \beta_\mathbf{p}^* \beta_\mathbf{q}^*} {\alpha _\mathbf{p}^* \alpha _\mathbf{q}^*} | \uparrow  \downarrow_\mathbf{p}, \uparrow_{-\mathbf{p}} \rangle^{\mathrm{out}}| \uparrow  \downarrow_\mathbf{q}, \downarrow_{-\mathbf{q}} \rangle ^{\mathrm{out}} \\
& + \sqrt {1 - \varepsilon^2} \alpha_\mathbf{p} \alpha_\mathbf{q} | \downarrow_\mathbf{p},0_{-\mathbf{p}}\rangle^{\mathrm{out}}| \uparrow _\mathbf{q},0_{-\mathbf{q}}\rangle^{\mathrm{out}} \\
& - \sqrt {1 - \varepsilon ^2} \frac{\alpha_\mathbf{p} \alpha _\mathbf{q} \beta_\mathbf{q}^*} {\alpha_\mathbf{q}^*} | \downarrow _\mathbf{p},0_{-\mathbf{p}}\rangle^{\mathrm{out}}| \uparrow \downarrow_\mathbf{q}, \uparrow_{-\mathbf{q}}\rangle^{\mathrm{out}} \\
& + \sqrt {1 - \varepsilon ^2} \frac{\alpha_\mathbf{q} \alpha_\mathbf{p} \beta_\mathbf{p}^*} {\alpha_\mathbf{p}^*}| \uparrow \downarrow _\mathbf{p}, \downarrow_{-\mathbf{p}}\rangle^{\mathrm{out}}| \uparrow_\mathbf{q},0_{-\mathbf{q}}\rangle^{\mathrm{out}} \\
& - \sqrt {1 - \varepsilon ^2} \frac{ \alpha _\mathbf{p} \alpha _\mathbf{q} \beta _\mathbf{p}^* \beta _\mathbf{q}^*} {\alpha _\mathbf{p}^* \alpha _\mathbf{q}^*} | \uparrow  \downarrow_\mathbf{p}, \downarrow_{-\mathbf{p}}\rangle^{\mathrm{out}}| \uparrow  \downarrow_\mathbf{q}, \uparrow _{-\mathbf{q}}\rangle^{\mathrm{out}}.
\end{array}
\end{equation}
In the following, we consider various pairs of out modes.

\subsection{$\rho _{\mathbf{p}, \mathbf{q}}  $  }

By tracing over the field modes $(-\mathbf{p},-\mathbf{q})$,  we obtain  the   reduced density matrix of the out modes $(\mathbf{p},\mathbf{q})$,  $\rho _{\mathbf{p},\mathbf{q}}  =  {\mathrm{Tr}}_{-\mathbf{p},-\mathbf{q}}(| \Phi \rangle^{\mathrm{in}} \langle \Phi |)$,
\begin{equation}\begin{array}{rl}
\rho_{\mathbf{p},\mathbf{q}}=& \varepsilon ^2 | \alpha_\mathbf{p}|^2 | \alpha_\mathbf{q}|^2 | \uparrow_\mathbf{p}, \downarrow_\mathbf{q} \rangle \langle \uparrow_\mathbf{p},\downarrow_\mathbf{q} | + \varepsilon \sqrt {1 - \varepsilon ^2} | \alpha_\mathbf{p} |^2 | \alpha_\mathbf{q} |^2 | \uparrow_\mathbf{p}, \downarrow_\mathbf{q} \rangle \langle  \downarrow_\mathbf{p}, \uparrow_\mathbf{q} | \\
& + \varepsilon ^2 | \alpha_\mathbf{p} |^2 | \beta_\mathbf{q} |^2 | \uparrow_\mathbf{p}, \uparrow \downarrow_\mathbf{q} \rangle \langle \uparrow_\mathbf{p}, \uparrow  \downarrow_\mathbf{q} |  + \varepsilon ^2 | \alpha_\mathbf{q} |^2 | \beta_\mathbf{p} |^2 | \uparrow  \downarrow_\mathbf{p}, \downarrow_\mathbf{q} \rangle  \langle   \uparrow   \downarrow_\mathbf{p}, \downarrow_\mathbf{q} | \\
& + | \beta_\mathbf{p} |^2 | \beta_\mathbf{q} |^2 | \uparrow   \downarrow_\mathbf{p}, \uparrow  \downarrow_\mathbf{q} \rangle \langle \uparrow   \downarrow_\mathbf{p}, \uparrow  \downarrow_\mathbf{q} | + (1 - \varepsilon ^2) | \alpha_\mathbf{p}  |^2 | \alpha _\mathbf{q} |^2 | \downarrow_\mathbf{p}, \uparrow_\mathbf{q} \rangle  \langle \downarrow_\mathbf{p}, \uparrow_\mathbf{q} | \\
&+ \varepsilon \sqrt {1 - \varepsilon ^2}  | \alpha_\mathbf{p} |^2 | \alpha_\mathbf{q} |^2 | \downarrow_\mathbf{p}, \uparrow _\mathbf{q} \rangle  \langle  \uparrow_\mathbf{p}, \downarrow_\mathbf{q} |   \\
& + (1 -\varepsilon ^2) | \alpha_\mathbf{p}  |^2 | \beta_\mathbf{q} |^2 | \downarrow_\mathbf{p}, \uparrow  \downarrow_\mathbf{q} \rangle \langle  \downarrow_\mathbf{p}, \uparrow \downarrow_\mathbf{q} |   \\
& + (1 -  \varepsilon ^2 ) | \alpha _\mathbf{q} |^2 | \beta _\mathbf{p} |^2 | \uparrow  \downarrow _\mathbf{p} ,  \uparrow _\mathbf{q}  \rangle  \langle \uparrow  \downarrow _\mathbf{p} ,  \uparrow _\mathbf{q} |,
\end{array}\end{equation}
which is a $7\times7$ matrix. Its eigenvalues are found to be $ 0 $, $|\alpha_\mathbf{p}|^2 |\alpha_\mathbf{q} |^2$, $\varepsilon ^2 | \alpha_\mathbf{p} |^2 | \beta_\mathbf{q} |^2$, $\varepsilon ^2 | \alpha_\mathbf{q} |^2 | \beta_\mathbf{p} |^2$, $| \beta_\mathbf{p} |^2 | \beta_\mathbf{q} |^2$, $ (1 -  \varepsilon ^2 ) | \alpha_\mathbf{p} |^2 | \beta_\mathbf{q} |^2$, $(1 - \varepsilon ^2 ) | \alpha_\mathbf{q} |^2 | \beta_\mathbf{p} |^2 $.

Tracing out the field mode $\mathbf{q}$ in  $\rho_{\mathbf{p},\mathbf{q}}$ yields $\rho_\mathbf{p} = {\mathrm{Tr}}_\mathbf{q}(\rho_{\mathbf{p},\mathbf{q}})$, which is
\begin{equation}   \rho_\mathbf{p} = \varepsilon ^2 | \alpha_\mathbf{p} |^2 | \uparrow_\mathbf{p} \rangle  \langle  \uparrow_\mathbf{p}  | + (1 - \varepsilon ^2 ) | \alpha_\mathbf{p} |^2  | \downarrow_\mathbf{p}  \rangle \langle  \downarrow_\mathbf{p} | +  | \beta_\mathbf{p} |^2 | \uparrow   \downarrow_\mathbf{p} \rangle  \langle \uparrow  \downarrow_\mathbf{p} |,  \end{equation}
with eigenvalues  $\varepsilon ^2 | \alpha_\mathbf{p} |^2$, $(1 - \varepsilon ^2) | \alpha_\mathbf{p}  |^2 $, $ | \beta_\mathbf{p} |^2$.

Tracing out the field mode $\mathbf{p}$ in $\rho_{\mathbf{p},\mathbf{q}}$ yields $\rho_\mathbf{q} = {\mathrm{Tr}}_\mathbf{p}(\rho_{\mathbf{p},\mathbf{q}})$, which is
\begin{equation}  \rho_\mathbf{q} = \varepsilon ^2 | \alpha_\mathbf{q} |^2 | \downarrow_\mathbf{q} \rangle  \langle \downarrow_\mathbf{q} | + (1 - \varepsilon ^2) |  \alpha_\mathbf{q} |^2  | \uparrow_\mathbf{q} \rangle \langle  \uparrow _\mathbf{q} | +  | \beta _\mathbf{q}  |^2 |  \uparrow   \downarrow _\mathbf{q} \rangle \langle   \uparrow   \downarrow _\mathbf{q} |,    \end{equation}
with eigenvalues $\varepsilon^2 | \alpha_\mathbf{q} |^2$, $(1 - \varepsilon ^2 ) | \alpha_\mathbf{q} |^2$, $ | \beta _\mathbf{q} |^2$.

Now we make use of the mutual information $I(\rho_{\mathbf{p},\mathbf{q}}) \equiv S(\rho_\mathbf{p}) + S(\rho_\mathbf{q}) -S(\rho_{\mathbf{p},\mathbf{q}}) $  to learn about the total correlation in  $\rho_{\mathbf{p},\mathbf{q}}$.  It is found to be
\begin{equation}  I(\rho_{\mathbf{p},\mathbf{q}}) = 2 S(\varepsilon ) | \alpha _\mathbf{p} |^2 | \alpha _\mathbf{q} |^2],  \end{equation}
where
\begin{equation}
S(\varepsilon )\equiv -\varepsilon ^2 \log _2 \varepsilon ^2  + (1 - \varepsilon ^2 ) \log _2 (1 -  \varepsilon ^2 )\end{equation} is the von Neumann entropy of the initial state
(\ref{initialent}).

For a given initial state in the absence of electric field, $I(\rho_{\mathbf{p},\mathbf{q}})$ reaches the  maximum $S(\varepsilon )$ when there is no electric field, i.e. $E_0=0$, which implies  $| \alpha _\mathbf{p} | = | \alpha _\mathbf{q} |=1$. It can reach the absolute maximum $2$ if the inital state in the absence of the electric field is maximally entangled, i.e. $\varepsilon ^2=1/2$, thus $S(\varepsilon) =1$. $I(\rho_{\mathbf{p},\mathbf{q}})$   monotonically decreases with  increase of $E_0$.

We have also studied the pairwise entanglement  in   $\rho_{\mathbf{p},\mathbf{q}}$, as quantified by its logarithmic negativity. For this purpose,  the partial transpose of $\rho_{\mathbf{p},\mathbf{q}}$ on mode $ \mathbf{p} $ is obtained as \begin{equation} \begin{array}{rl}
\rho_{\mathbf{p},\mathbf{q}}^{T_A}=& \varepsilon ^2 | \alpha_\mathbf{p}|^2 | \alpha_\mathbf{q}|^2 | \uparrow_\mathbf{p}, \downarrow_\mathbf{q} \rangle \langle \uparrow_\mathbf{p},\downarrow_\mathbf{q} | + \varepsilon \sqrt {1 - \varepsilon ^2} | \alpha_\mathbf{p} |^2 | \alpha_\mathbf{q} |^2 | \downarrow_\mathbf{p}, \downarrow_\mathbf{q} \rangle \langle  \uparrow_\mathbf{p}, \uparrow_\mathbf{q} | \\
& + \varepsilon ^2 | \alpha_\mathbf{p} |^2 | \beta_\mathbf{q} |^2 | \uparrow_\mathbf{p}, \uparrow \downarrow_\mathbf{q} \rangle \langle \uparrow_\mathbf{p}, \uparrow  \downarrow_\mathbf{q} |  + \varepsilon ^2 | \alpha_\mathbf{q} |^2 | \beta_\mathbf{p} |^2 | \uparrow  \downarrow_\mathbf{p}, \downarrow_\mathbf{q} \rangle  \langle   \uparrow   \downarrow_\mathbf{p}, \downarrow_\mathbf{q} | \\
& + | \beta_\mathbf{p} |^2 | \beta_\mathbf{q} |^2 | \uparrow   \downarrow_\mathbf{p}, \uparrow  \downarrow_\mathbf{q} \rangle \langle \uparrow   \downarrow_\mathbf{p}, \uparrow  \downarrow_\mathbf{q} | + (1 - \varepsilon ^2) | \alpha_\mathbf{p} |^2 | \alpha _\mathbf{q} |^2 | \downarrow_\mathbf{p}, \uparrow_\mathbf{q} \rangle  \langle \downarrow_\mathbf{p}, \uparrow_\mathbf{q} |  \\
&+ \varepsilon \sqrt {1 - \varepsilon ^2}  | \alpha_\mathbf{p} |^2 | \alpha_\mathbf{q} |^2 | \uparrow_\mathbf{p}, \uparrow _\mathbf{q} \rangle  \langle  \downarrow_\mathbf{p}, \downarrow_\mathbf{q} |   \\
& + (1 -\varepsilon ^2) | \alpha_\mathbf{p}|^2 | \beta_\mathbf{q} |^2 | \downarrow_\mathbf{p}, \uparrow  \downarrow_\mathbf{q} \rangle \langle  \downarrow_\mathbf{p}, \uparrow \downarrow_\mathbf{q} |   \\
& + (1 - \varepsilon ^2)|\alpha _\mathbf{q} |^2 |  \beta _\mathbf{p} |^2 |  \uparrow   \downarrow _\mathbf{p} ,  \uparrow _\mathbf{q}  \rangle  \langle   \uparrow  \downarrow _\mathbf{p} ,  \uparrow _\mathbf{q} |,
\end{array}
\end{equation}
of which the eigenvalues   are $ \varepsilon^2 | \alpha_\mathbf{p} |^2 | \alpha_\mathbf{q} |^2 $, $\pm \varepsilon \sqrt {1 - \varepsilon^2 } | \alpha_\mathbf{p} |^2 | \alpha_\mathbf{q} |^2 $, $ \varepsilon^2  | \alpha_\mathbf{p} |^2 | \beta_\mathbf{q} |^2 $, $\varepsilon^2 | \alpha_\mathbf{q} |^2 | \beta_\mathbf{p} |^2 $, $| \beta_\mathbf{p} |^2 | \beta_\mathbf{q} |^2 $, $(1 - \varepsilon ^2 ) | \alpha_\mathbf{p} |^2 | \alpha_\mathbf{q} |^2 $, $(1 -  \varepsilon^2 ) | \alpha_\mathbf{p} |^2 | \beta_\mathbf{q} |^2 $, $ (1 -  \varepsilon^2 ) | \alpha_\mathbf{q} |^2 | \beta_\mathbf{p} |^2 $.
Hence the logarithmic negativity is found to be
\begin{equation}  N( \rho _{\mathbf{p},\mathbf{q}} ) =  \log _2 (1 + 2\varepsilon \sqrt {1 -  \varepsilon ^2 }  | \alpha _\mathbf{p} |^2 | \alpha _\mathbf{q} |^2 ),   \end{equation}
which reaches its maximum $ \log _2 (1 + 2\varepsilon \sqrt {1 -  \varepsilon ^2 }) $ when $| \alpha _\mathbf{p} | = | \alpha _\mathbf{q} |=1$, as satisfied when $E_0=0$, and monotonically decreases with $E_0$. Again, the absolute maximum $1$ can be reached if the initial state in absence of electric field is maximally entangled, i.e.  $\varepsilon = 1/\sqrt{2}$.

\begin{figure}
\centering
\scalebox{0.45}{\includegraphics{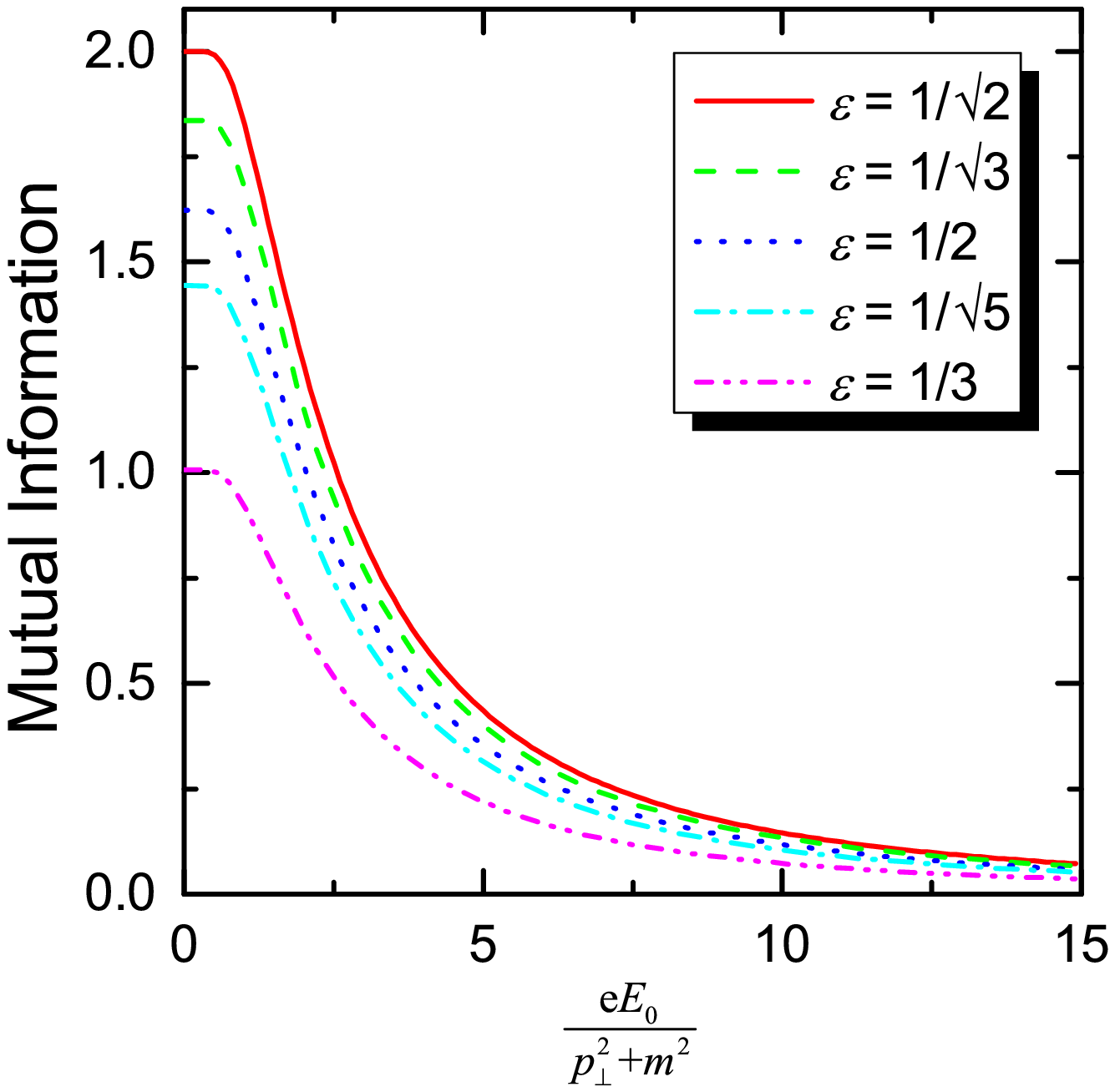}}\scalebox{0.45}{\includegraphics{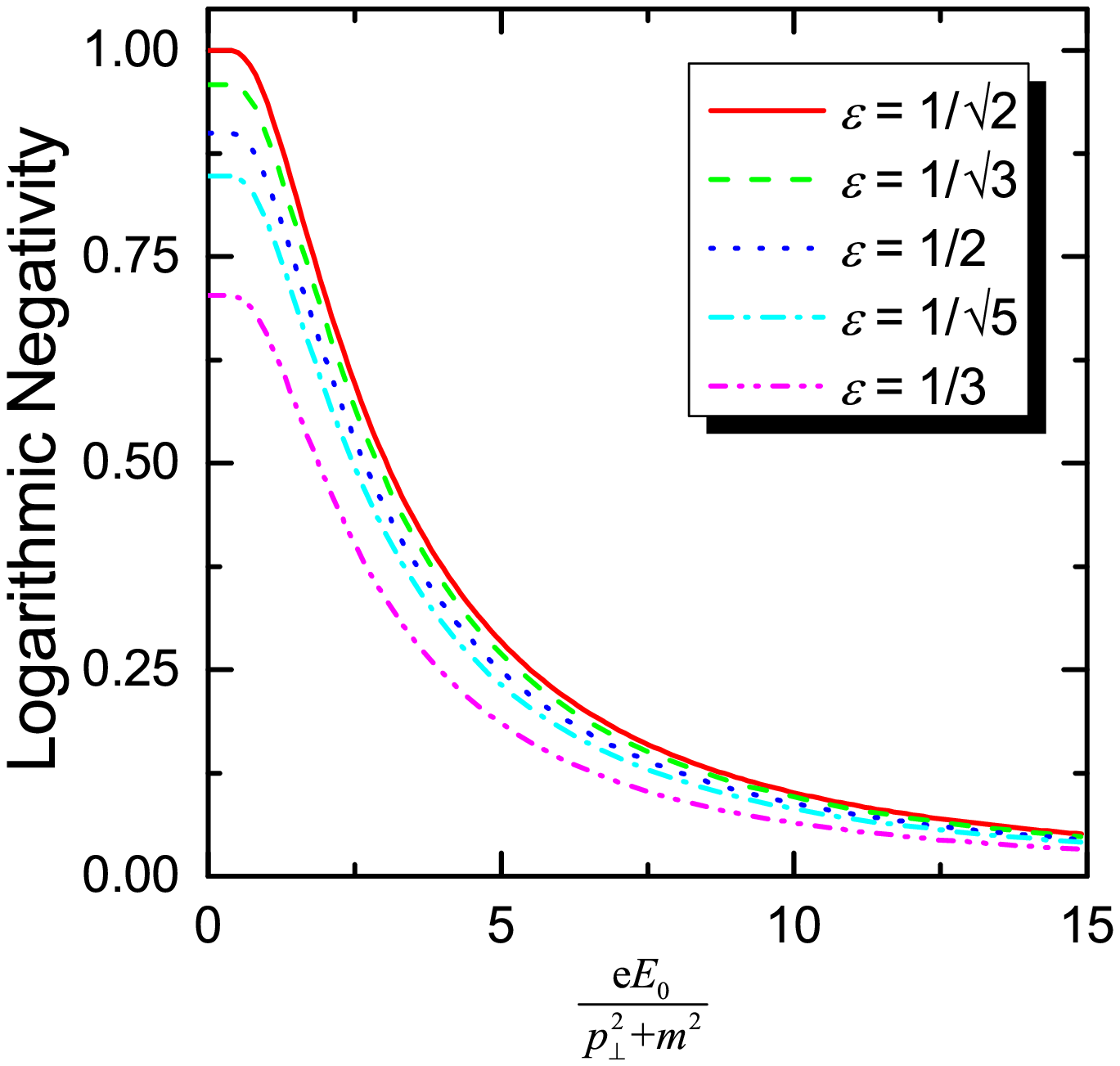}}
\caption{ The mutual information $I(\rho_{\mathbf{p},\mathbf{q}})$ and logarithmic negativity $N(\rho_{\mathbf{p},\mathbf{q}})$ as functions of the dimensionless parameter $\frac{eE_0}{p_\perp^2+m^2}$, where  $E_0$  is the strength of the constant electric field.  It is assumed  that    $ p_\bot ^2=q_\bot ^2$. \label{fig2} }
\end{figure}

The dependence of both mutual information $I(\rho_{\mathbf{p},\mathbf{q}})$ and logarithmic negativity $N(\rho_{\mathbf{p},\mathbf{q}})$ on electric field $E_0$ is shown in Fig.~\ref{fig2}.  Both quantities monotonically decrease and asymptotically approach   zero with the increase of the electric field $E_0$. Moreover, for $\varepsilon \leq 1/\sqrt{2}$, the closer to $1/\sqrt{2}$ $\varepsilon$, i.e. the larger the initial entanglement, the larger   $I(\rho_{\mathbf{p},\mathbf{q}})$ and  $N(\rho_{\mathbf{p},\mathbf{q}})$.

Initially the entanglement is just that between modes  $\mathbf{p}$ and $\mathbf{q}$. Roughly speaking, the electric field causes the entanglement to be redistributed to be between other pairs of modes. The correlation is mostly contributed by the entanglement.

\subsection{$\rho _{\mathbf{p}, - \mathbf{q}}$ and $\rho _{-\mathbf{p},\mathbf{q}}$ }

Tracing out the field modes $(-\mathbf{p},\mathbf{q})$, we obtain $\rho _{\mathbf{p}, - \mathbf{q}}  = {\mathrm{Tr}} _{ - \mathbf{p},\mathbf{q}} ( | \Phi \rangle _ {\mathrm{in}} \langle \Phi  |)$.  The mutual information is obtained as
\begin{equation}  I(\rho _{\mathbf{p},-\mathbf{q}} ) = - 2S(\varepsilon ) | \alpha _\mathbf{p} |^2 | \beta _\mathbf{q} |^2,  \end{equation}
while the logarithmic negativity is
\begin{equation} N(\rho _{\mathbf{p}, -\mathbf{q}}) = \log _2(1 + 2\varepsilon \sqrt {1 - \varepsilon ^2} | \alpha _\mathbf{p}|^2 | \beta _\mathbf{q} |^2 ).   \end{equation}

\begin{figure}
\centering
\scalebox{0.45}{\includegraphics{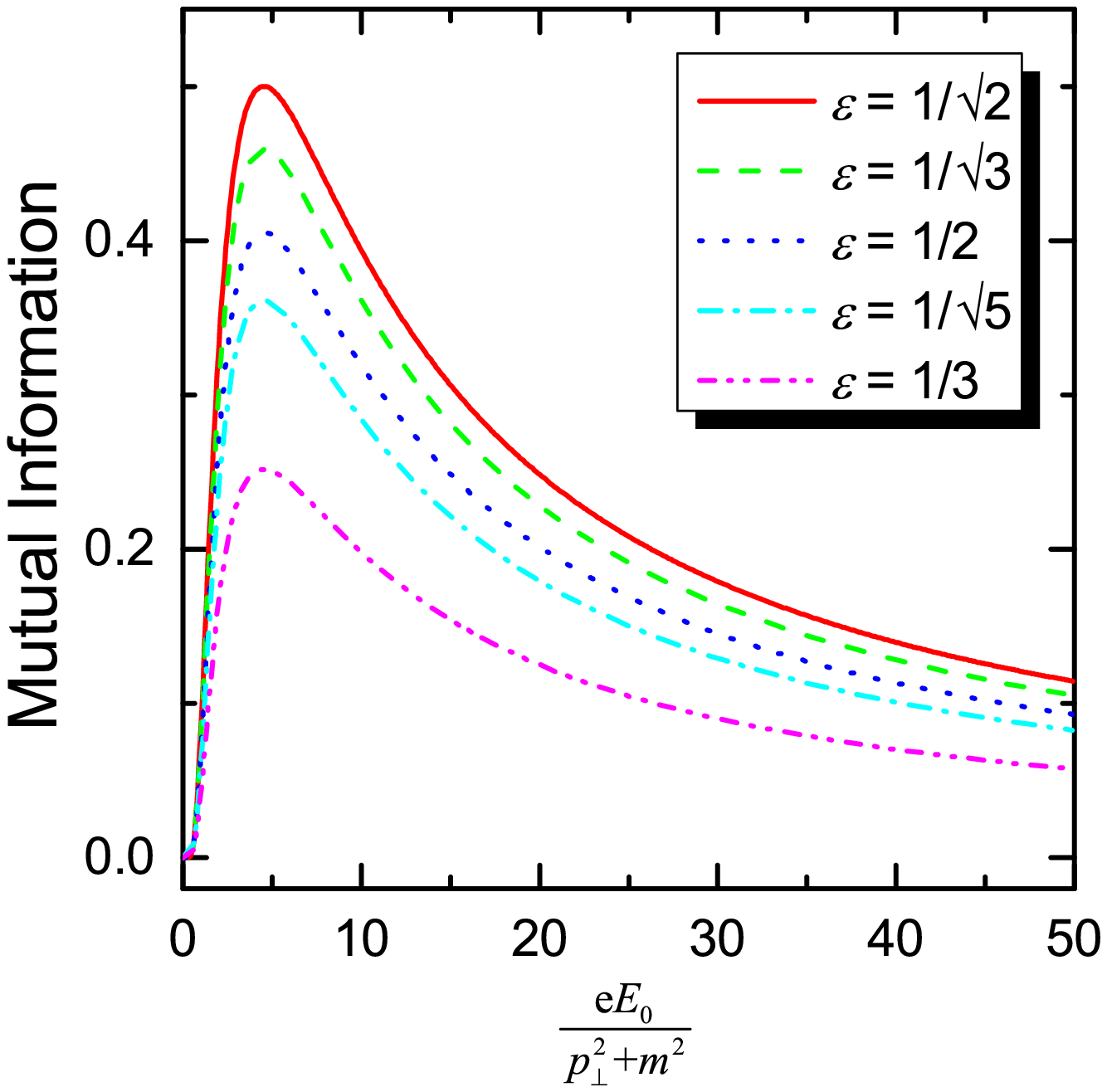}}\scalebox{0.45}{\includegraphics{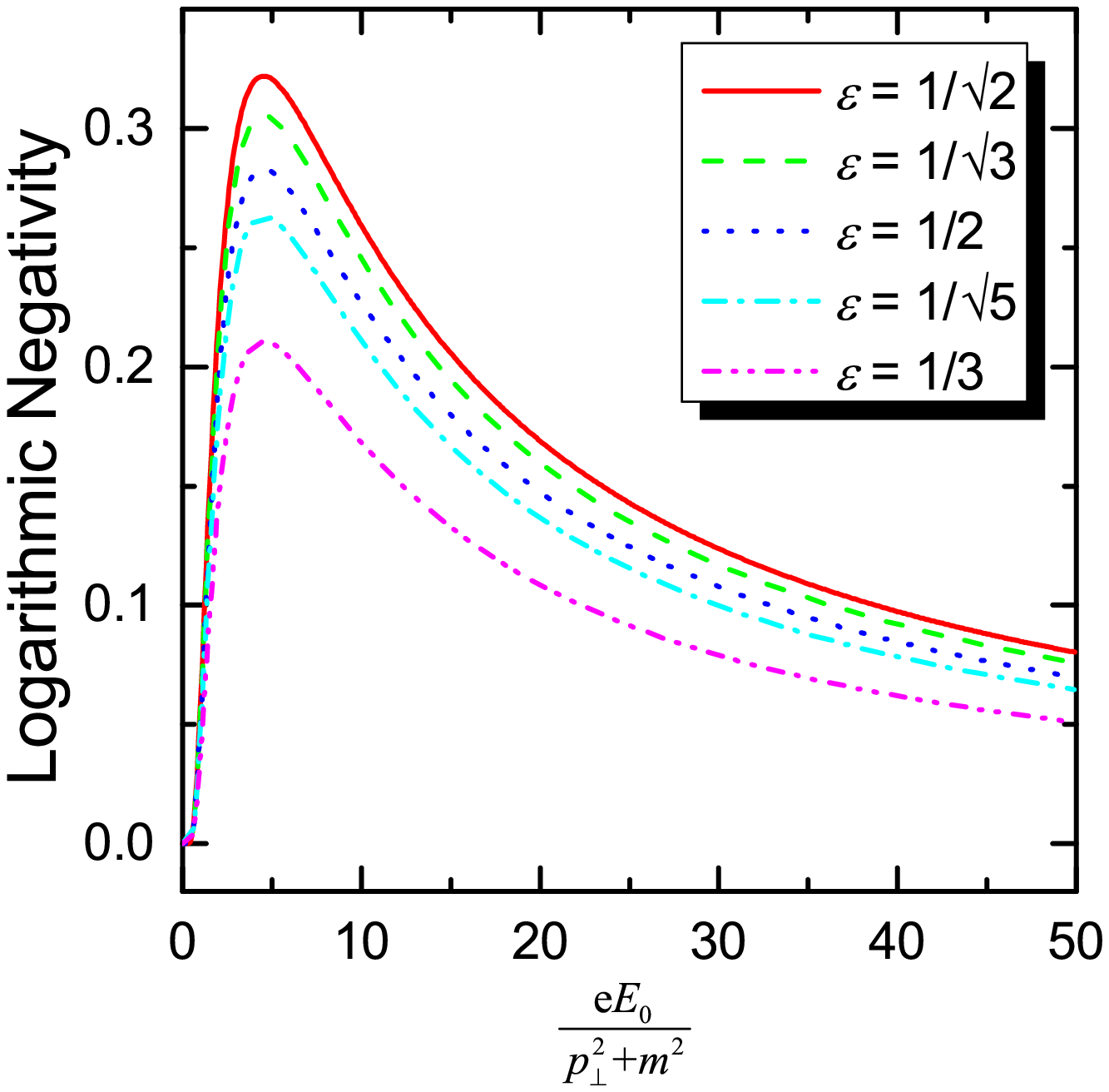}}
\caption{ The mutual information $I(\rho_{\mathbf{p},-\mathbf{q}})$ and logarithmic negativity $N(\rho_{\mathbf{p},-\mathbf{q}})$ as functions of the dimensionless parameter $\frac{eE_0}{p_\perp^2+m^2}$, where  $E_0$  is the strength of the constant electric field.  It is assumed  that    $ p_\bot ^2=q_\bot ^2$. \label{fig3} }
\end{figure}

As shown in Fig.~\ref{fig3}, in the absence of electric field, both the mutual information $I(\rho_{\mathbf{p},-\mathbf{q}})$ and the logarithmic negativity $N(\rho_{\mathbf{p},-\mathbf{q}})$ are zero. With the increase of the electric field, they  increase to maxima at a same value $E_{c}$ of $E_0$ where $ | \alpha _\mathbf{p} |^2 | \beta _\mathbf{q} |^2 $ is maximized,  then decreases to zero asymptotically. For simplicity, we consider the case $ p_\bot ^2=q_\bot ^2$, which means $\alpha_\mathbf{p}= \alpha_\mathbf{q}$ and $\beta_\mathbf{p}= \beta_\mathbf{q}$. Then it can be seen that
$I(\rho_{\mathbf{p},-\mathbf{q}})$ and  $N(\rho_{\mathbf{p},-\mathbf{q}})$ reach their respective  maxima $S(\varepsilon )/2 \leq 1/2$ and $\log _2(1 + \varepsilon \sqrt {1 - \varepsilon ^2} |/2  ) \leq \log_2(5/4) $   when $|\alpha_\mathbf{p}|^2=  |\beta_\mathbf{q}|^2 =1/2$,  as satisfied by $E_c= \frac{\pi(m^2 + p_\bot ^2)}{e \ln 2 }$.

Similarly, tracing out the field modes $\mathbf{p}$ and $-\mathbf{q}$, we obtain  $\rho _{-\mathbf{p}, \mathbf{q}}  = { \mathrm{Tr} } _{ \mathbf{p},-\mathbf{q}} ( | \Phi \rangle _ {\mathrm{in}} \langle \Phi |)$.  The mutual information is obtained as
\begin{equation}  I(\rho _{-\mathbf{p},\mathbf{q}} )= - 2S(\varepsilon ) | \alpha _\mathbf{q} |^2 | \beta _\mathbf{p} |^2, \end{equation}
while the logarithmic negativity is
\begin{equation} N(\rho _{-\mathbf{p}, \mathbf{q}}) = \log _2(1 + 2\varepsilon \sqrt {1 - \varepsilon ^2} | \alpha _\mathbf{q} |^2 | \beta _\mathbf{p} |^2 ).   \end{equation}

$\rho_{-\mathbf{p},\mathbf{q}}$ can be obtained from $\rho_{\mathbf{p},-\mathbf{q}}$ simply by exchanging the generic subscripts $\mathbf{p}$ and $\mathbf{q}$ as well as $\varepsilon$ and $\sqrt{1-\varepsilon^2}$. Therefore, the properties of the mutual information and logarithmic negativity of   $\rho_{-\mathbf{p},\mathbf{q}}$ are the same as those of $\rho_{\mathbf{p},-\mathbf{q}}$ with  $\varepsilon$ replaced as $\sqrt{1-\varepsilon^2}$.

\subsection{$\rho _{-\mathbf{p}, -\mathbf{q}}$  }

Tracing out the field modes $\mathbf{p}$ and $\mathbf{q}$, we obtain  $\rho _{-\mathbf{p}, -\mathbf{q}}  = { \mathrm{Tr} } _{ \mathbf{p}, \mathbf{q}} ( | \Phi \rangle _ {\mathrm{in}} \langle \Phi  |)$. It is found that
\begin{equation}  I(\rho _{-\mathbf{p},-\mathbf{q}} ) = - 2S(\varepsilon ) | \beta _\mathbf{p} |^2 | \beta _\mathbf{q} |^2,  \end{equation}
and
\begin{equation} N(\rho _{-\mathbf{p},-\mathbf{q}}) = \log _2(1 + 2\varepsilon \sqrt {1 - \varepsilon ^2} | \beta _\mathbf{p} |^2 | \beta _\mathbf{q} |^2 ).   \end{equation}

\begin{figure}[htb]
\centering
\scalebox{0.45}{\includegraphics{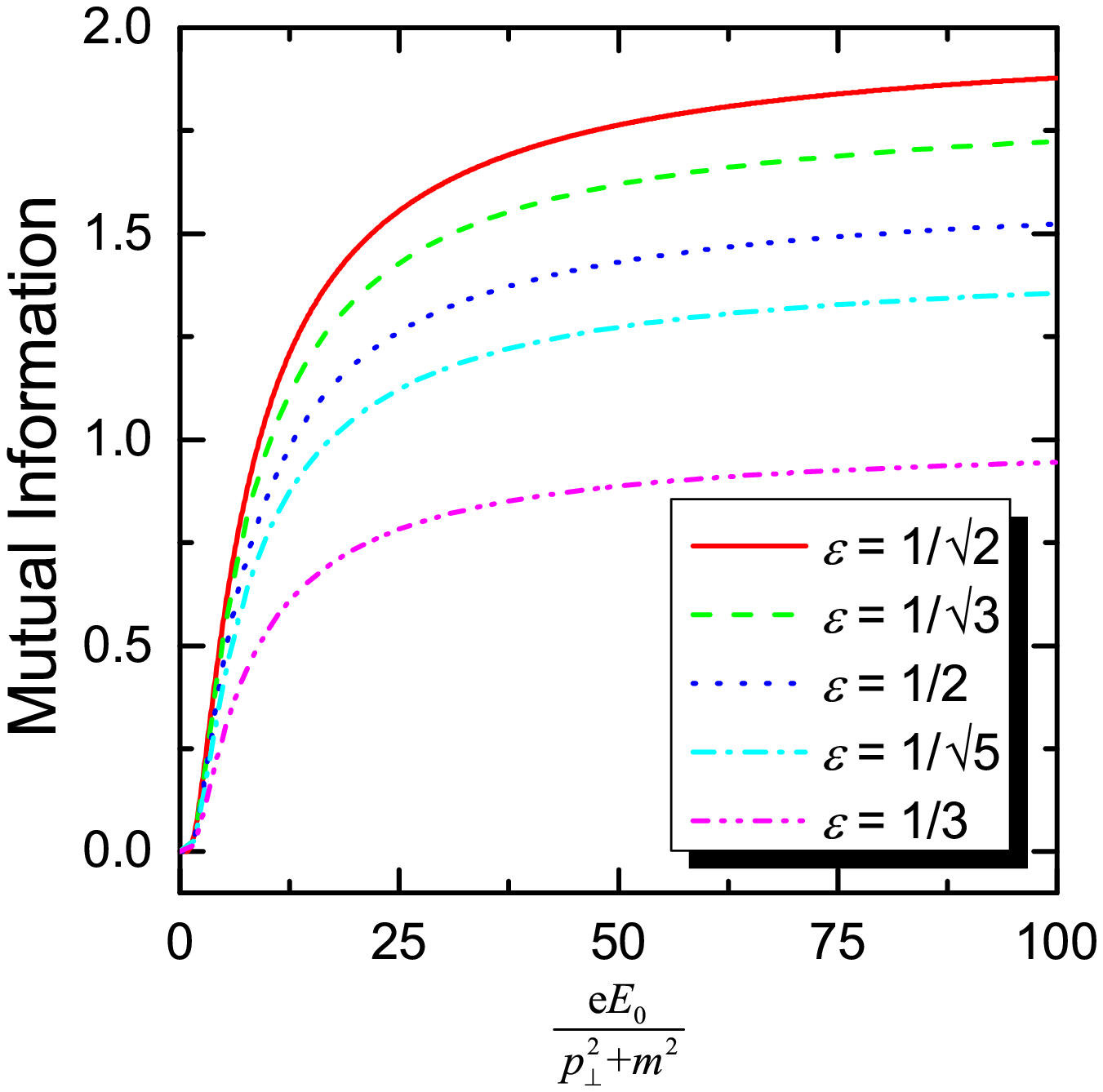}}\scalebox{0.45}{\includegraphics{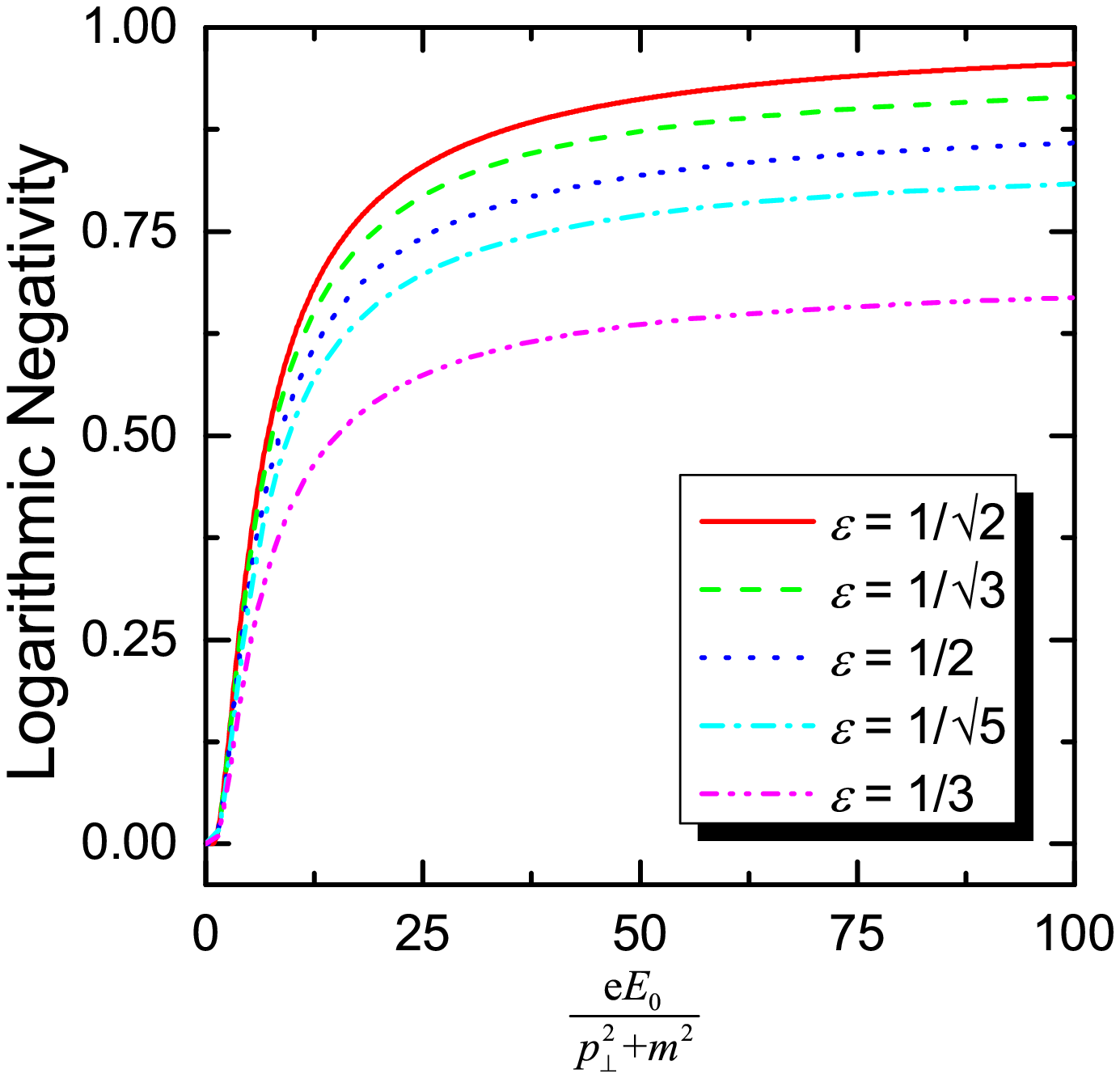}}
\caption{ The mutual information $I(\rho_{-\mathbf{p},-\mathbf{q}})$ and logarithmic negativity $N(\rho_{-\mathbf{p},-\mathbf{q}})$ as functions of the dimensionless parameter $\frac{eE_0}{p_\perp^2+m^2}$, where  $E_0$  is the strength of the constant electric field.  It is assumed  that    $ p_\bot ^2=q_\bot ^2$. \label{fig4} }
\end{figure}

Fig.~\ref{fig4} shows how the mutual information and the logarithmic negativity in  $\rho _{-\mathbf{p}, -\mathbf{q}}$ depend on the strength $E_0$ of the electric field. They are zero at $E_0=0$ and monotonically increase with $E_0$.  The trend is opposite to  that of the $\rho_{\mathbf{p}, \mathbf{q}}$. Note that in the absence of the electric field, there is no entanglement or correlation between modes $-\mathbf{p}$ and $-\mathbf{q}$.

\subsection{$\rho _{\mathbf{p}, - \mathbf{p}}$ and $\rho _{\mathbf{q},-\mathbf{q}}$}

Tracing out the field modes $\mathbf{q}$ and $-\mathbf{q}$, one obtains $\rho _{\mathbf{p}, -\mathbf{p}}  = { \mathrm{Tr} } _{ \mathbf{q}, -\mathbf{q}} ( | \Phi \rangle _ {\mathrm{in}} \langle \Phi  |)$. It is found that
\begin{equation} I(\rho _{\mathbf{p}, - \mathbf{p}}) =  - 2(| \alpha _\mathbf{p} |^2 \log _2 | \alpha _\mathbf{p} |^2  +  | \beta _\mathbf{p} |^2 \log _2 | \beta _\mathbf{p} |^2 ), \end{equation}
and
\begin{equation}  N( \rho _{\mathbf{p}, - \mathbf{p}} ) = \log _2 [1 + 2(1 - 2 \varepsilon ^2  + 2 \varepsilon ^4 ) | \alpha _\mathbf{p} |^2 | \beta _\mathbf{p} |^2 ] .  \end{equation}
It can be seen that both $I(\rho _{\mathbf{p}, - \mathbf{p}})$ and $N( \rho _{\mathbf{p}, - \mathbf{p}} )$ reach the maxima at $| \alpha _\mathbf{p} |^2 =| \beta _\mathbf{p} |^2=1/2$, i.e. at a finite value of $E_0$.

\begin{figure}
\centering
\scalebox{0.45}{\includegraphics{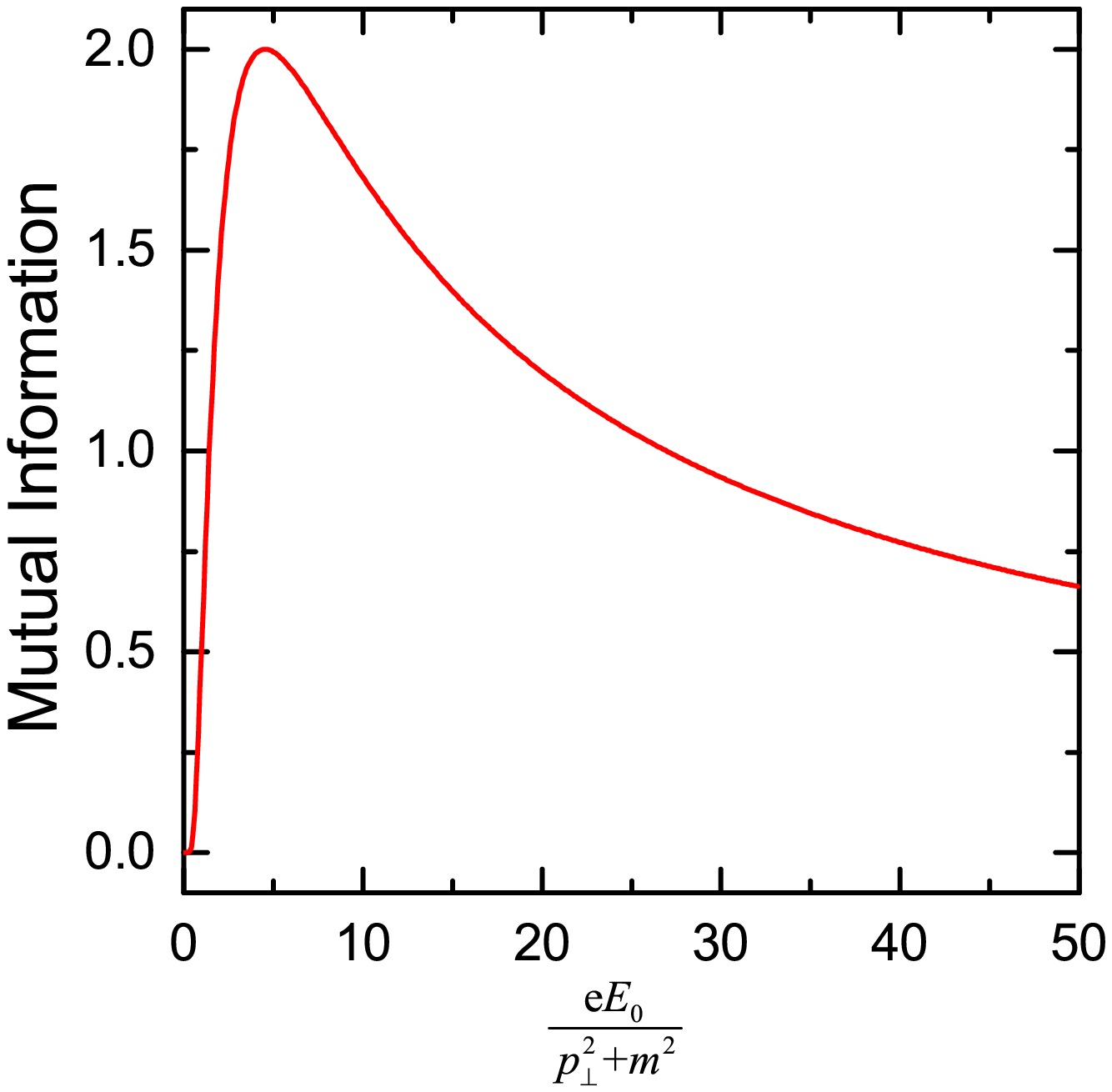}}\scalebox{0.45}{\includegraphics{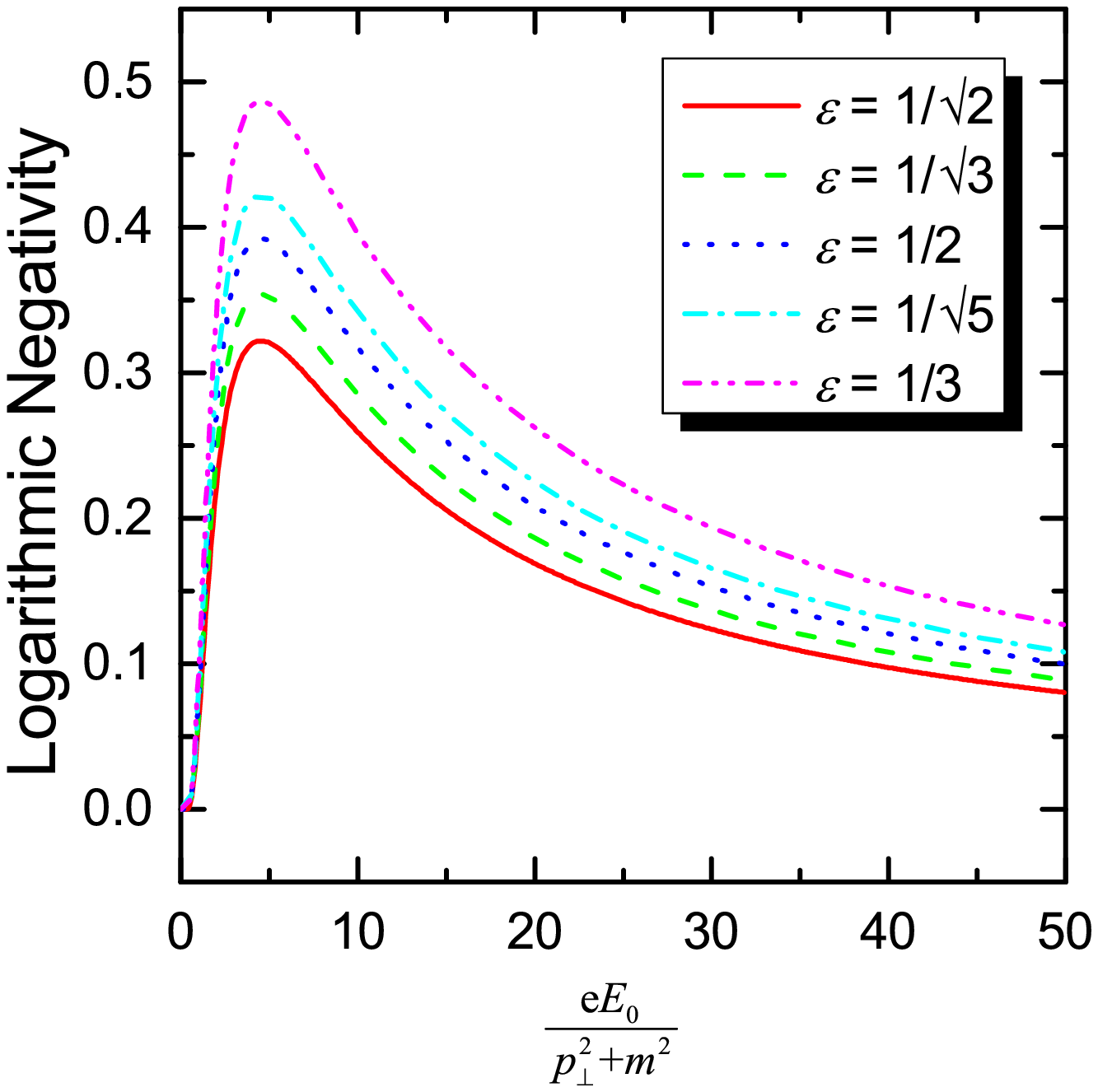}}
\caption{ The mutual information $I(\rho_{\mathbf{p},-\mathbf{p} })$ and logarithmic negativity $N(\rho_{\mathbf{p},-\mathbf{p} })$ as functions of the dimensionless parameter $\frac{eE_0}{p_\perp^2+m^2}$, where  $E_0$  is the strength of the constant electric field.  It is assumed  that    $ p_\bot ^2=q_\bot ^2$.  \label{fig5} }
\end{figure}

Note that the mutual information $I(\rho_{\mathbf{p},-\mathbf{p}})$ is independent of   $\varepsilon $. In fact it comes only from the Schwinger production of particle-antiparticle pairs, since $\mathbf{p}$and $-\mathbf{p}$ are connected by Bogoliubov coefficients.

As shown in Fig.~\ref{fig5},  it   increases to the maximum $2$ at  $E_0=\frac{\pi(m^2 + p_\bot ^2)}{e\ln 2 }$, then decreases and  approaches zero asymptotically. With the increase of  $E_0$, the logarithmic negativity $N(\rho_{\mathbf{p},-\mathbf{p}})$ increases from zero to the maximum $ \log _2 [1 + (1 - 2 \varepsilon ^2  + 2 \varepsilon ^4 ) /2] $   at  $E_0=\frac{\pi(m^2 + p_\bot ^2)}{e\ln 2 }$, and then decreases to zero asymptotically. The closer to $1/\sqrt{2}$ the coefficient $\varepsilon$, the larger the initial entanglement between $\mathbf{p}$ and $-\mathbf{p}$, the less the logarithmic negativity $N(\rho_{\mathbf{p},-\mathbf{p}})$,   in contrast with the logarithmic negativities   $N(\rho_{\mathbf{p},\mathbf{q}})$,   $N(\rho_{\mathbf{p},-\mathbf{q}})$, $N(\rho_{-\mathbf{p},\mathbf{q}})$ and $N(\rho_{-\mathbf{p},-\mathbf{q}})$.

Similarly, tracing out the field modes $\mathbf{p}$ and $-\mathbf{p}$ yields $\rho _{\mathbf{q}, -\mathbf{q}}  = { \mathrm{Tr} } _{\mathbf{p}, -\mathbf{p}} ( | \Phi \rangle _ {\mathrm{in}} \langle \Phi  |)$. We obtain
\begin{equation} I(\rho _{\mathbf{q}, -\mathbf{q}}) =  - 2(| \alpha _\mathbf{q} |^2 \log _2 | \alpha _\mathbf{q} |^2  +  | \beta _\mathbf{q} |^2 \log _2 | \beta _\mathbf{q} |^2 ),   \end{equation}
and
\begin{equation}  N( \rho _{\mathbf{q}, -\mathbf{q}} ) = \log _2 [1 + 2(1 - 2 \varepsilon ^2  + 2 \varepsilon ^4 ) | \alpha _\mathbf{q} |^2 | \beta _\mathbf{q} |^2 ].  \end{equation}
which show similar  behavior with   $I(\rho _{\mathbf{p}, -\mathbf{p}})$ and $N(\rho _{\mathbf{p}, -\mathbf{p}})$, as $\mathbf{q}$ and $\mathbf{p}$ are only momentum labels.

\subsection{General Relations }

For $\rho_{\mathbf{p},\mathbf{q}}$, $\rho _{\mathbf{p}, -\mathbf{q}}$, $\rho _{ - \mathbf{p},\mathbf{q}}$, and $\rho _{ -\mathbf{p}, - \mathbf{q}}$, the mutual information and the logarithmic negativity are both determined by the product of the modular square of two Bogoliubov coefficients. For the  subscripts of the density matrices,  particle modes $\mathbf{p}$ and $\mathbf{q}$ correspond to $\alpha_\mathbf{p}$ or $\alpha_\mathbf{q}$, antiparticle modes   $-\mathbf{p}$ and  $-\mathbf{q}$ correspond to $\beta_\mathbf{p}$ or $\beta_\mathbf{q}$. For each of these four density matrices $\rho$, the mutual information $I(\rho)$ and the logarithmic negativity $N(\rho)$ are related simply as
\begin{equation}  I(\rho )  =  \frac{S(\varepsilon )}{ \varepsilon \sqrt{1- \varepsilon ^2}}(2^{N(\rho)}-1).
\end{equation}
$I(\rho)$  and $N(\rho)$ are synchronized, peaking and vanishing simultaneously.

We also find  the following two identities
\begin{eqnarray}
 I(\rho _{\mathbf{p},\mathbf{q}}) + I(\rho _{\mathbf{p}, -\mathbf{q}}) + I(\rho _{ - \mathbf{p},\mathbf{q}}) + I(\rho _{ -\mathbf{p}, - \mathbf{q}})&  =  &- 2S(\varepsilon ), \label{identity1} \\
2^{N( \rho _{\mathbf{p},\mathbf{q}} )}  +  2^{N( \rho _{\mathbf{p}, -\mathbf{q}} )}  +  2^{N( \rho _{ - \mathbf{p},\mathbf{q}} )} + 2^{N( \rho _{ -\mathbf{p}, - \mathbf{q}} )}  & = &4 + 2\varepsilon \sqrt {1 -  \varepsilon ^2}.  \label{identity2}
\end{eqnarray}
In (\ref{identity1}), $- 2S(\varepsilon )$ is the total mutual information of the state in the absence of the electric field, coming only from $\rho _{\mathbf{p},\mathbf{q}}$.
In (\ref{identity1}), in the absence of the electric field, the first term on LHS contributes $1+ 2\varepsilon \sqrt {1 -  \varepsilon ^2}$, while each of the rest three terms on LHS contributes 1, i.e. no entanglement.

Thus these two identities represent two conservation laws.  They suggest that some form of the combination of the four kinds of mutual information and that of the four kinds of entanglement are conserved, and are redistributed by the  electric field.

In the case that the electric field is only applied to the particle of momentum $\mathbf{p}$, we have $ | \alpha_\mathbf{q} |^2  = 1 $ , $ | \beta _\mathbf{q} |^2 = 0 $. Thus $I(\rho _{\mathbf{p},-\mathbf{q}}) = I(\rho _{ -\mathbf{p}, - \mathbf{q}}) =  N(\rho _{ \mathbf{p},-\mathbf{q}}) = N(\rho _{ -\mathbf{p}, - \mathbf{q}}) = 0$.   Consequently,
\begin{eqnarray}
I( \rho _{\mathbf{p},\mathbf{q}} ) + I( \rho _{ - \mathbf{p},\mathbf{q}} ) &= &- 2S(\varepsilon ),  \\
 2^{N( \rho _{\mathbf{p},\mathbf{q}} )}  +  2^{N( \rho _{ -\mathbf{p},\mathbf{q}} )} & =& 2 + 2\varepsilon \sqrt {1 -  \varepsilon ^2 }.  \end{eqnarray}

Similar discussions can be made for the entangled states  of the forms of  $\varepsilon | \uparrow _\mathbf{p},0_{ - \mathbf{p}} \rangle ^{ \mathrm{in}} | \downarrow _\mathbf{q},0_{ - \mathbf{q}} \rangle ^{ \mathrm{in}}  - \sqrt {1 - \varepsilon ^2}  | \downarrow _\mathbf{p},0_{ - \mathbf{p}} \rangle^{ \mathrm{in}} | \uparrow _\mathbf{q},0_{ -\mathbf{q}} \rangle^{\mathrm{in}}$ and $ \varepsilon | \uparrow _\mathbf{p},0_{ - \mathbf{p}} \rangle ^{ \mathrm{in}} | \uparrow_\mathbf{q},0_{ - \mathbf{q}} \rangle ^{ \mathrm{in}}  \pm \sqrt {1 - \varepsilon ^2}  | \downarrow _\mathbf{p},0_{ - \mathbf{p}} \rangle^{ \mathrm{in}} | \downarrow _\mathbf{q},0_{ - \mathbf{q}} \rangle^{\mathrm{in}}$. If $ p_\bot ^2=q_\bot ^2$, the results are the same as here for $\varepsilon | \uparrow _\mathbf{p},0_{ - \mathbf{p}} \rangle ^{ \mathrm{in}} | \downarrow _\mathbf{q},0_{ - \mathbf{q}} \rangle ^{ \mathrm{in}}  + \sqrt {1 - \varepsilon ^2}  | \downarrow _\mathbf{p},0_{ - \mathbf{p}} \rangle^{ \mathrm{in}} | \uparrow _\mathbf{q},0_{ -\mathbf{q}} \rangle^{\mathrm{in}}$.

\section{ A pulsed electric field \label{pulse}  }

Now  we  investigate the influence of a  pulsed electric field, which is assumed to be  along the $z$ direction and is of the Sauter-type  $E_0 \mathrm{sech}^2(t /\tau)$, where $\tau$ is the width of  the pulsed electric field~\cite{sauter}. Therefore
\begin{equation}
A_\mu = \left( 0,0,0, -E_0\tau \tanh \left(\frac{t}{\tau }\right) \right).   \end{equation}
By solving the two linearly independent asymptotic solutions of the Dirac equation in the regimes $t^{\mathrm{in}} = - \infty $ and $t^{\mathrm{out}} = + \infty $ respectively, one  finds the Bogoliubov coefficients~\cite{kim,Ebadi}
\begin{equation}  | \alpha_\mathbf{k} |^2  = \frac{ \cosh [ \pi \tau ( \omega _{ \mathrm{out}} +  \omega ^{ \mathrm{in}} ) ] - \cosh ( 2 \pi \lambda )} {2 \sinh ( \pi \tau  \omega ^{ \mathrm{in}} )\sinh \ ( \pi \tau \omega ^{\mathrm{out}})}  ,   \label{alpha} \end{equation}
\begin{equation}  | \beta_\mathbf{k} |^2  = \frac{ \cosh ( 2 \pi \lambda ) - \cosh [ \pi \tau ( \omega _{ \mathrm{out}} - \omega ^{ \mathrm{in}} ) ] } {2 \sinh ( \pi \tau  \omega ^{ \mathrm{in}} )\sinh \ ( \pi \tau \omega ^{\mathrm{out}})}  ,  \label{beta}  \end{equation}
where
\begin{equation}  \lambda  = e E_0 \tau ^2 ,   \end{equation}
\begin{equation}\omega^{\mathrm{in}}=\sqrt{(k_z + e E_0 \tau )^2 + k_\bot ^2 + m^2} ,
\end{equation}
\begin{equation}\omega^{\mathrm{out}}=\sqrt{(k_z - e E_0 \tau )^2 + k_\bot ^2 + m^2} .
\end{equation}
$E(t) \to 0$ as $\tau  \to 0$, $E(t) \to E_0$ as  $\tau  \to  + \infty $.

Note that the  formulae in Sections \ref{const} and \ref{ent}, including  those for the mutual information and logarithmic negativity,  still apply, but with $| \alpha_\mathbf{k} |^2$ and   $| \beta_\mathbf{k} |^2 $ now given in Eqs.~(\ref{alpha}) and (\ref{beta}).  Hence the mutual information and logarithmic negativity depend on not only $E_0$ but also $\tau$. The general trend  is in consistency with the case of constant field, but now with new features because there are two parameters.

\begin{figure}[htb]
\centering
\includegraphics[width=0.5\textwidth]{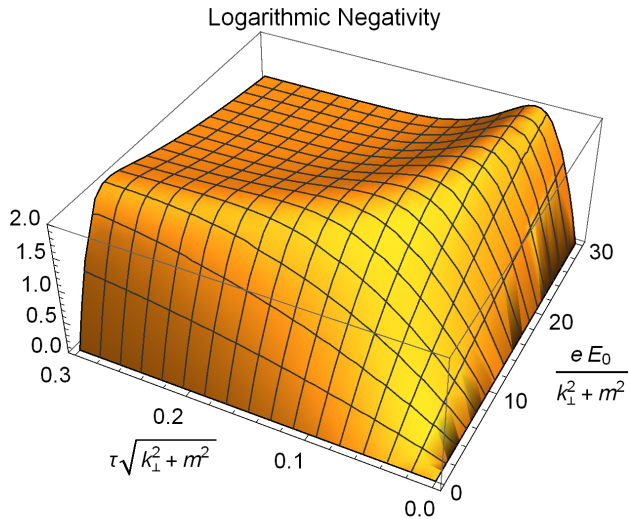}
\caption{ The logarithmic negativity $N( \rho_{\mathbf{k}, - \mathbf{k}} )$  as functions of   the dimensionless parameters $\frac{eE_0}{k_\perp^2+m^2 }$ and $\tau \sqrt{k_\perp^2+m^2 }$.   It is assumed  that   $k_z= \sqrt{k_\perp^2+m^2 }$. \label{fig6} }
\end{figure}

In parallel with the above study on constant field, we first consider the entanglement created by the pulsed field from the vacuum, as given in (\ref{vacuum}). Fig.~\ref{fig6} shows the dependence of the logarithmic negativity $ N( \rho_{\mathbf{k}, - \mathbf{k}} )$   on the dimensionless parameters  $1/\mu\equiv eE_0/(k_\perp^2+m^2)$ and  $\tau\sqrt{ k_\perp^2+m^2}$ of the pulsed electric field.  For any given value of $\tau$,  with the increase of  $E_0$,  $N( \rho_{\mathbf{k}, - \mathbf{k}} )$ increases to the maximum $2$ and then decreases asymptotically to zero.  The smaller $\tau$, the larger the value of $E_0$ where   $N( \rho_{\mathbf{k}, - \mathbf{k}} )$ is maximal.  As $\tau \rightarrow \infty$,   the maximum  $2$ of $N( \rho_{\mathbf{k}, - \mathbf{k}} )$ is reached   at  $E_0= \frac{\pi(k_\bot ^2+m^2)}{e\ln 2 }$, recovering the case of the constant electric field. For a given value of $E_0 \leq \frac{ k_\bot ^2+\pi(m^2 )}{e\ln 2 }$, $N( \rho_{\mathbf{k}, - \mathbf{k}} )$ increases monotonically with the increase of $\tau$, and asymptotically approaches   a certain value $\leq 2$.  For a given value of $E_0 > \frac{\pi(k_\perp^2+m^2)}{e \ln 2 }$,   with the increase of $\tau$,  $N( \rho_{\mathbf{k}, - \mathbf{k}} )$ increases up to the maximum $2$ and then decreases, and  approaches asymptotically  a certain value  dependent on $E_0 $. As $\tau \rightarrow \infty$, the dependence of $N( \rho_{\mathbf{k}, - \mathbf{k}} )$  on $E_0$ is just as shown in Fig.~\ref{fig1}.

Now we consider two initially entangled particles acted on by a pulsed electric field. The initial state is just $ |\Phi_{\mathbf{p},\mathbf{q},-\mathbf{p},-\mathbf{q}}\rangle ^{\mathrm{in}}$ as given in    (\ref{initialent}).  The calculation is similar to that in the previous section.

\begin{figure}
\centering
\scalebox{0.6}{\includegraphics{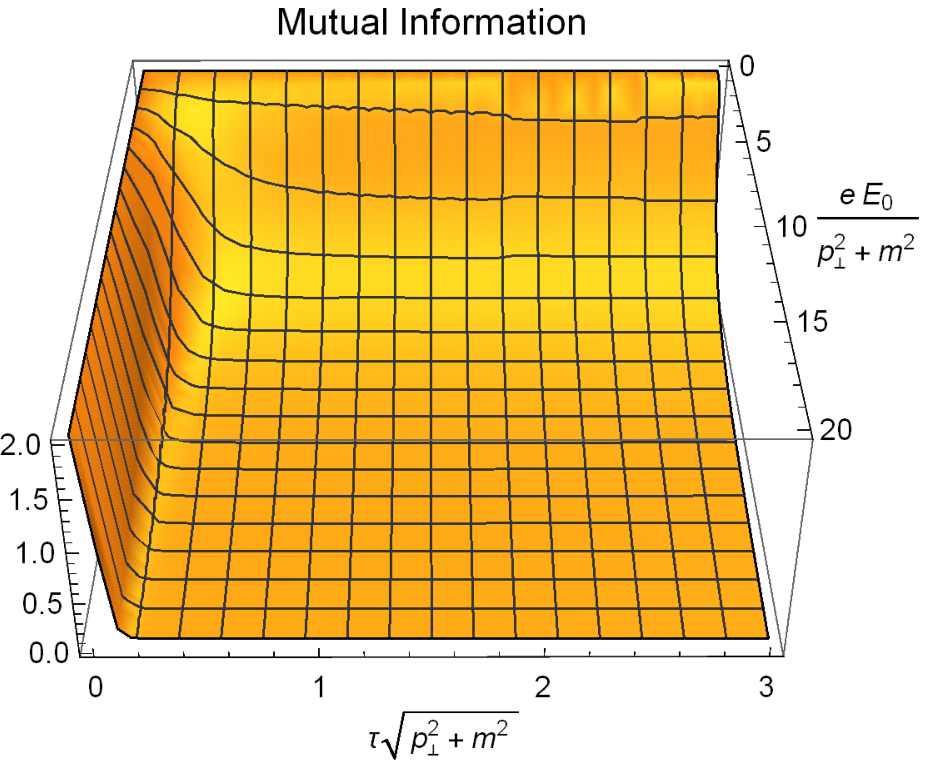}}\scalebox{0.63}{\includegraphics{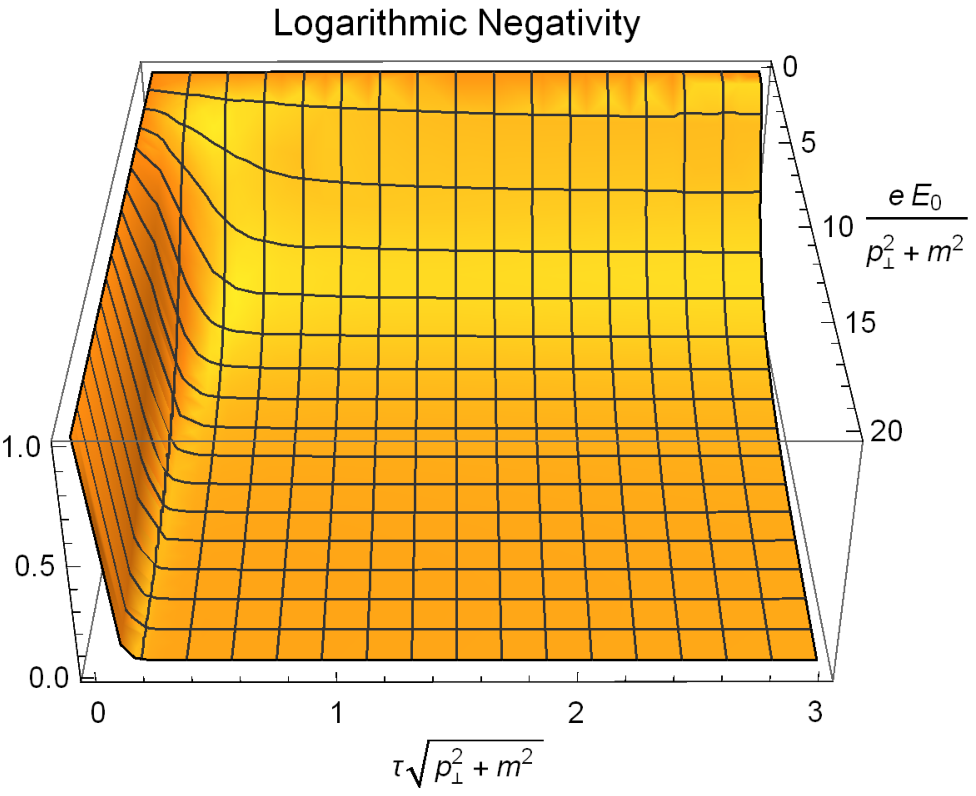}}
\caption{ The mutual information $I(\rho_{\mathbf{p},\mathbf{q}})$ and logarithmic negativity $N(\rho_{\mathbf{p},\mathbf{q}})$  as functions of   the dimensionless parameters $\frac{eE_0}{p_\perp^2+m^2}$ and $\tau \sqrt{p_\perp^2+m^2 }$.   It is assumed  that    $p_z=q_z= \sqrt{p_\perp^2+m^2 }=\sqrt{q_\perp^2+m^2 }$,  $\varepsilon=1/ \sqrt{2}$. \label{fig7} }
\end{figure}

Fig.~\ref{fig7}  shows how the pulsed electric field affects the mutual information and the logarithmic negativity between modes $\mathbf{p}$ and $\mathbf{q}$. For any value of $\tau$,   with the increase of $E_0$, $I(\rho_{\mathbf{p},\mathbf{q}})$ and $N(\rho_{\mathbf{p},\mathbf{q}})$ monotonically decrease and asymptotically approach  zero. The larger $\tau$, the quicker they decrease when $E_0$ is small. For a fixed value of $E_0$, with the increase of $\tau$,  $I(\rho_{\mathbf{p},\mathbf{q}})$ and $N(\rho_{\mathbf{p},\mathbf{q}})$ monotonically decrease and asymptotically approach  certain values dependent on $E_0$. The larger $E_0$, the quicker they decrease when  $\tau$ is small.  As $\tau \rightarrow \infty$, the dependence of $N( \rho_{\mathbf{p},  \mathbf{q}} )$  on $E_0$ is just as shown in Fig.~\ref{fig2}.

\begin{figure}
\centering
\scalebox{0.6}{\includegraphics{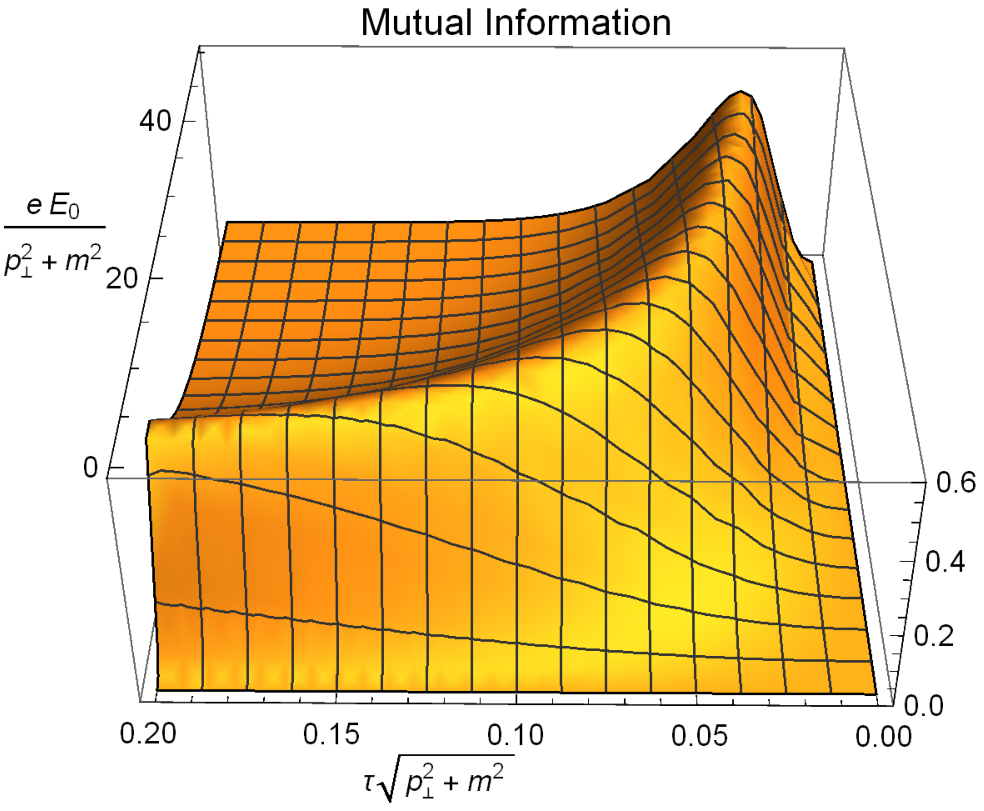}}\scalebox{0.63}{\includegraphics{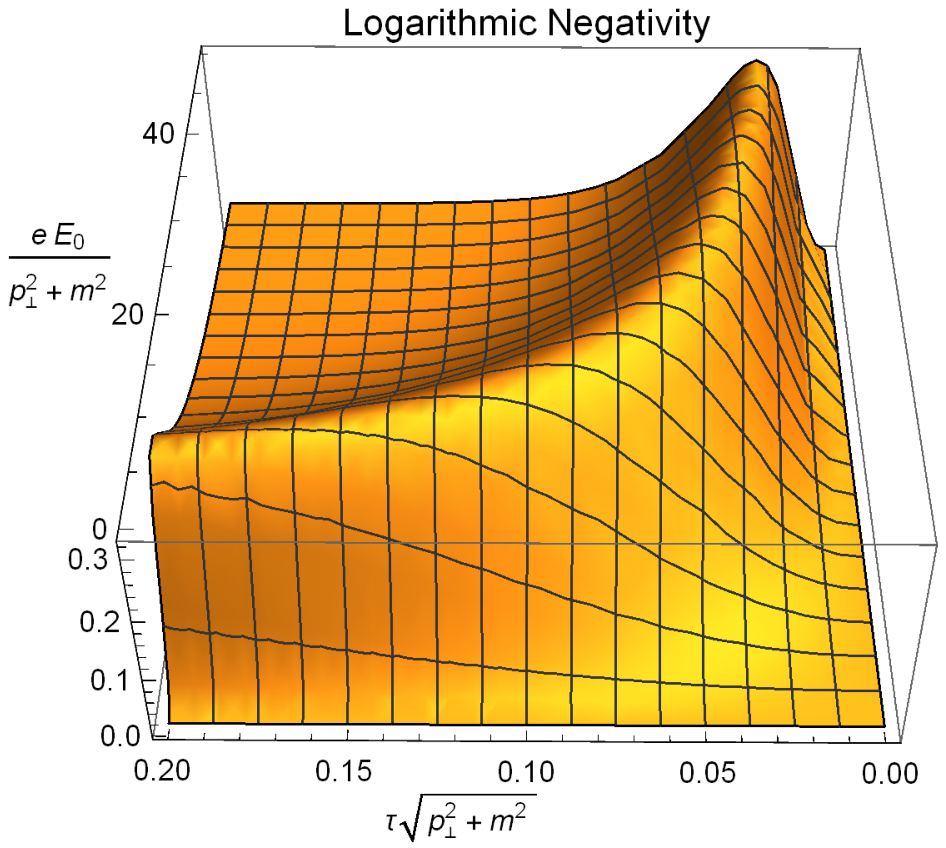}}
\caption{ The mutual information $I(\rho_{\mathbf{p},-\mathbf{q}})$ and logarithmic negativity $N(\rho_{\mathbf{p},-\mathbf{q}})$   as functions of   the dimensionless parameters $\frac{eE_0}{p_\perp^2+m^2}$ and $\tau \sqrt{p_\perp^2+m^2 }$.   It is assumed  that    $p_z=q_z= \sqrt{p_\perp^2+m^2 }=\sqrt{q_\perp^2+m^2 }$, $\varepsilon=1/ \sqrt{2}$.   Under these parameter values, the maxima of $I(\rho_{\mathbf{p},-\mathbf{q}})$ and $N(\rho_{\mathbf{p},-\mathbf{q}})$ are $\frac{1}{2}$ and $\log_2(\frac{5}{4})$, respectively.  \label{fig8} }
\end{figure}

Fig.~\ref{fig8}  shows how the mutual information and logarithmic negativity of $\rho(\mathbf{p},-\mathbf{q})$ depend on the strength $E_0$ and the width $\tau$ of  the pulsed electric field.  For any given value of $\tau$,  with the increase of  $E_0$,   $I( \rho_{\mathbf{p}, - \mathbf{q}} )$ and  $N( \rho_{\mathbf{p}, - \mathbf{q}} )$ respectively increase to their  maxima at some value $E_{c}(\tau)$, where $|\alpha_\mathbf{p}|^2|\beta_\mathbf{q}|^2$ is maximized, and then decrease asymptotically to zero. $E_{c}(\tau)$ decreases with $\tau$.   For a given value of $E_0 \leq E_{c}(\infty)$, $I( \rho_{\mathbf{p}, - \mathbf{q}} )$ and  $N( \rho_{\mathbf{p}, - \mathbf{q}} )$   increase   monotonically with $\tau$, and asymptotically approaches    certain values $\leq 2$.  For a given value of $E_0 > E_{c}(\infty)$,   with the increase of $\tau$,  $I( \rho_{\mathbf{p}, - \mathbf{q}} )$ and $N( \rho_{\mathbf{k}, - \mathbf{k}} )$ increase up to the maxima    and then decrease and  approach  asymptotically  certain values  dependent on $E_0$, with the dependence at $\tau \rightarrow \infty$ as shown in Fig.~\ref{fig3}.

\begin{figure}
\centering
\scalebox{0.65}{\includegraphics{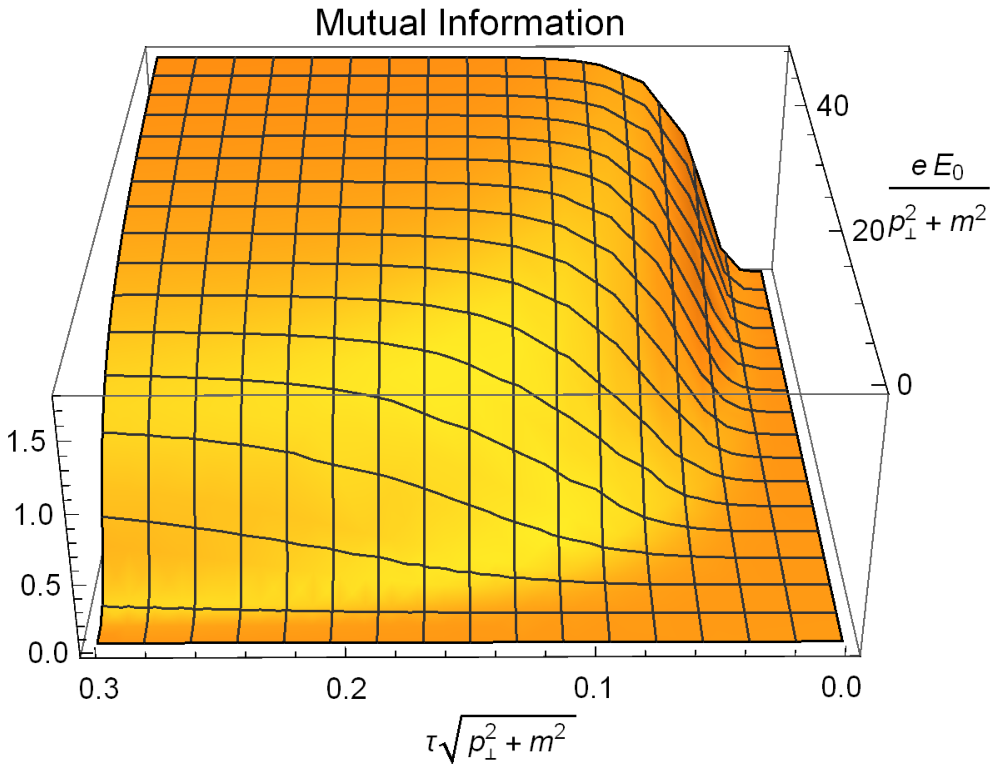}}\scalebox{0.63}{\includegraphics{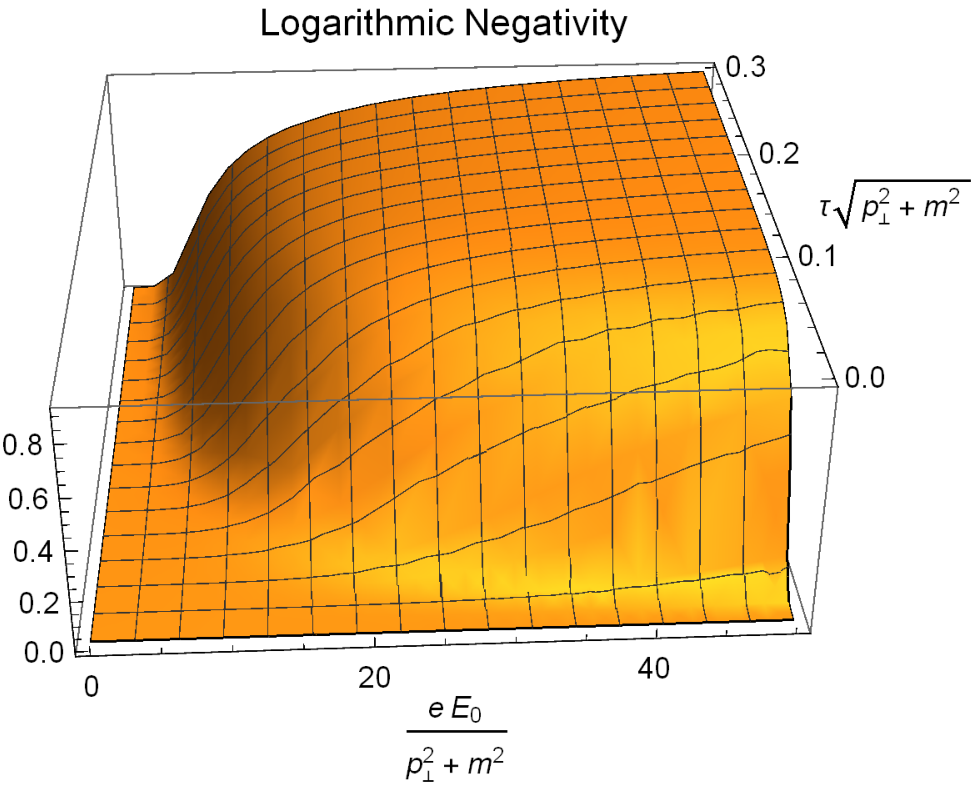}}
\caption{ The mutual information $I(\rho_{-\mathbf{p},-\mathbf{q}})$ and logarithmic negativity $N(\rho_{-\mathbf{p},-\mathbf{q}})$   as functions of   the dimensionless parameters $\frac{eE_0}{p_\perp^2+m^2}$ and $\tau \sqrt{p_\perp^2+m^2 }$.   It is assumed  that    $p_z=q_z= \sqrt{p_\perp^2+m^2 }=\sqrt{q_\perp^2+m^2 }$, $\varepsilon=1/ \sqrt{2}$.  \label{fig9} }
\end{figure}

Fig.~\ref{fig9}  shows how the mutual information and logarithmic negativity  of $\rho(-\mathbf{p},-\mathbf{q})$ depend on the strength $E_0$ and the width $\tau$ of  the pulsed electric field.  For any given value of $\tau$, with the increase of $E_0$, $I(\rho_{-\mathbf{p},-\mathbf{q}})$ and $N(\rho_{-\mathbf{p},-\mathbf{q}})$ monotonically increase and asymptotically approach  the maxima $2$ and $ 1$, respectively.  The larger $\tau$, the quicker they increase when $E_0$ is small.
For a given value of $E_0$, with the increase of $\tau$,
$I(\rho_{-\mathbf{p},-\mathbf{q}})$ and $N(\rho_{-\mathbf{p},-\mathbf{q}})$ monotonically increase and asymptotically approach certain values dependent on $E_0$.   The larger $E_0$, the quicker they increase when $\tau$ is small.
As $\tau \rightarrow \infty$,   the case of the constant electric field is recovered, as shown in Fig.~\ref{fig4}.

\begin{figure}
\centering
\scalebox{0.6}{\includegraphics{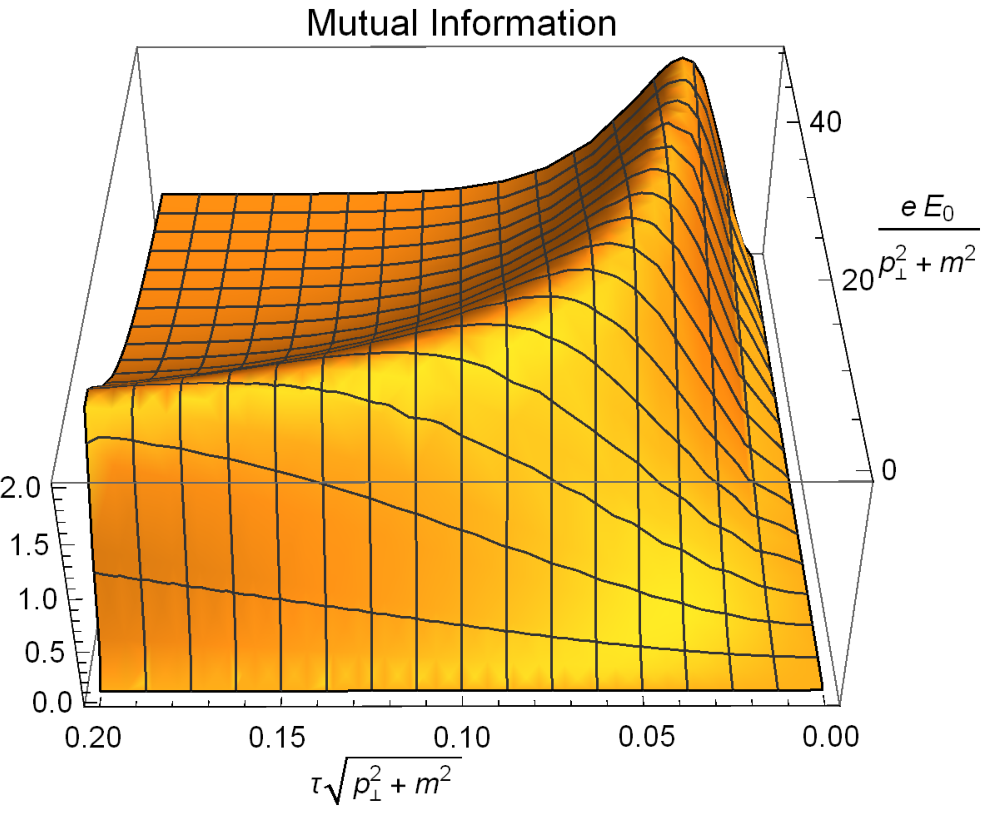}}\scalebox{0.63}{\includegraphics{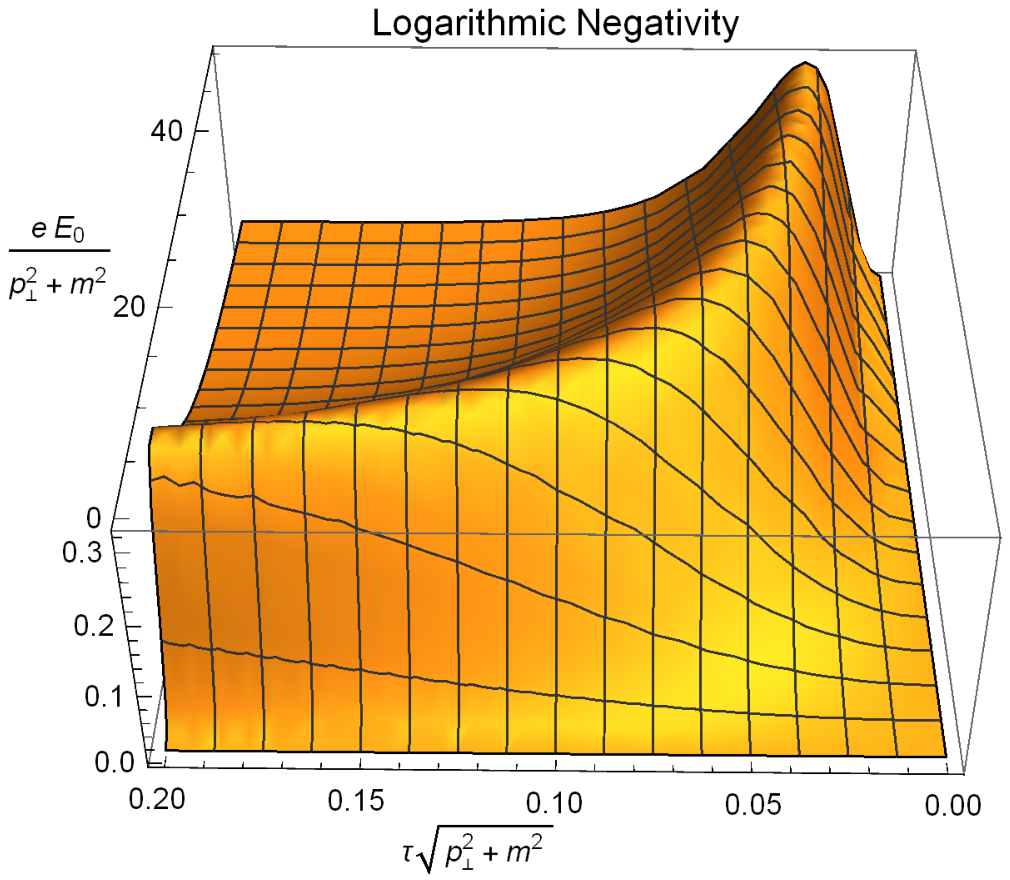}}
\caption{ The mutual information $I(\rho_{\mathbf{p},-\mathbf{p}})$ and logarithmic negativity $N(\rho_{\mathbf{p},-\mathbf{p}})$   as functions of   the dimensionless parameters $\frac{eE_0}{p_\perp^2+m^2}$ and $\tau \sqrt{p_\perp^2+m^2 }$.   It is assumed  that    $p_z= \sqrt{p_\perp^2+m^2 }$, $\varepsilon=1/ \sqrt{2}$.   Under these parameter values, the maxima of $I(\rho_{\mathbf{p},-\mathbf{p}})$ and  $N(\rho_{\mathbf{p},-\mathbf{p}})$ are $2$ and $\log_2(\frac{5}{4})$, respectively.  \label{fig10} }
\end{figure}

Fig.~\ref{fig10}  shows how the mutual information and logarithmic negativity  of $\rho(\mathbf{p},-\mathbf{p})$ depend on the strength $E_0$ and the width $\tau$ of  the pulsed electric field.  For any given value of $\tau$,  with the increase of  $E_0$,   $I( \rho_{\mathbf{p}, - \mathbf{p}} )$ and  $N( \rho_{\mathbf{p}, - \mathbf{p}} )$ respectively increase to their  maxima at certain values $E_{c}(\tau)$, where $| \alpha _\mathbf{p} |^2 =| \beta _\mathbf{p} |^2=1/2$, and then decrease asymptotically to zero.  $E_{c}(\tau)$ decreases with $\tau$, and   $E_{c}(\infty) = \frac{\pi(m^2 + p_\bot ^2)}{e \ln 2 }$. For a given value of $E_0 \leq E_{c}(\infty)$, $I( \rho_{\mathbf{p}, - \mathbf{p}} )$  and  $N( \rho_{\mathbf{p}, - \mathbf{p}} )$ increase   monotonically with $\tau$, and asymptotically approaches    certain values.  For a given value of $E_0 >E_{c}(\infty)$,   with the increase of $\tau$,  $I( \rho_{\mathbf{p}, - \mathbf{p}} )$  and   $N( \rho_{\mathbf{p}, - \mathbf{p}} )$ increase  up to the maxima  and then decrease, and  approach asymptotically  certain values  dependent on $E_0$.   As $\tau \rightarrow \infty$,   the case of the constant electric field is recovered, which is shown in Fig.~\ref{fig5}.

\section{Summary and discussions \label{summary}  }

In this paper, we have studied how the entanglement between different modes is changed by constant and pulsed electric fields.
We use the mutual information to quantify the total correlation and the logarithmic negativity to quantify the quantum entanglement.

First, we considered a constant electric field, which can induce the creation of particle-antiparticle pairs from the vacuum.  The  entanglement generated between modes $\mathbf{k}$ and $-\mathbf{k}$ is not monotonic with the electric field strength $E_0$. It first increases to the maximum, and then decreases asymptotically  towards zero.

We have also studied how an electric field affects the preexisting  entanglement. We choose an initial state to be entangled  between two particle modes $\mathbf{p}$ and $\mathbf{q}$.    We have calculated  the pairwise mutual information and logarithmic negativity  in the reduced density matrices   $\rho_{\mathbf{p},\mathbf{q}}$, $\rho_{\mathbf{p},-\mathbf{q}}$, $\rho_{-\mathbf{p},\mathbf{q}}$, $\rho_{-\mathbf{p},-\mathbf{q}}$, $\rho_{\mathbf{p},-\mathbf{p}}$, and $\rho_{\mathbf{q},-\mathbf{q}}$.

It is found that the mutual information and logarithmic negativity in the density matrix $\rho(\mathbf{p},\mathbf{q})$ of two particle modes monotonically decrease with the electric field strength $E_0$.  This is because they are determined by the product of the modular squares of $\alpha_\mathbf{p}$ and $\alpha_\mathbf{q}$, which both decrease with the increase of $E_0$.  Complementarily,  the mutual information and logarithmic negativity in the  density matrix $\rho(-\mathbf{p},-\mathbf{q})$ of two antiparticle modes   increase monotonically with   $E_0$.  This is because they are determined by $\beta_\mathbf{p}$ and $\beta_\mathbf{q}$.  Also bear in mind that the preexisting entanglment is   between the two particle modes  $\mathbf{p}$ and $\mathbf{q}$.

On the other hand,  in each of the four possible reduced density matrices of the particle-antiparticle modes $\rho_{\mathbf{p},-\mathbf{q}}$, $\rho_{-\mathbf{p},\mathbf{q}}$,  $\rho_{\mathbf{p},-\mathbf{p}}$, and $\rho_{\mathbf{q},-\mathbf{q}}$, the mutual information and logarithmic negativity first increase up to maxima, and then decrease asymptotically towards  zero. This is because they are determined by the product of the modular squares of  $\alpha_\mathbf{p}$ and $\beta_\mathbf{q}$.

We have also studied the effect of a pulsed electric field. The dependence on the Bogoliubov coefficients remain the same as the case of a constant field, but these coefficients    depend on  both the strength  $E_0$ and the decay width  $\tau$. The case of a constant field is recovered when   $\tau \rightarrow \infty$. For a density matrix of particle-antiparticle modes, at a given value of $\tau$, the mutual information and logarithmic negativity  maximize at a certain value of $E_0$ respectively. At a given value of $E_0$ no greater than the value corresponding to its maximum at $\tau\rightarrow\infty$,  the mutual information and logarithmic negativity increase monotonically with $\tau$, otherwise it reaches the maximum at   a certain value of $\tau$ and then decreases. For a density matrix of particle-particle modes, the mutual information and logarithmic negativity both monotonically decrease  with  both $E_0$ and $\tau$.  For a density matrix of antiparticle-antiparticle modes, the mutual information and logarithmic negativity both monotonically increase with   both $E_0$ and $\tau$.

We have obtained some relations and identities representing  conservation laws of various kinds of mode entanglement, which indicate the transfer of quantum information and are consistent with the calculations on these mode pairs. They are valid no matter whether the electric field is constant or time-dependent. Such information seems to be beyond the traditional quantities such as correlation functions and  distribution functions.

To conclude, we have studied how the electric field influences the pairwise mode entanglement through the Schwinger effect. It is demonstrated that quantum entanglement can be used as a theoretical tool to analyze Schwinger effect. The features of entanglement gives information on the Schwinger mechanism beyond that contained in the traditional quantities such as particle number distribution.  Hopefully the use of entanglement could be generalized to other processes in quantum electrodynamics, and in fundamental physics in general. For example, it would be interesting to see whether entanglement can shed new light on problems such as renormalization.

\end{document}